# Light-Induced Photomechanical Patterning of Ferroelectric Polarization

*Alban Degezelle[1*], Jonas Strobelt[2], Sarah Loebner[2], Moussa Mebarki[3], Stéphane Fusil[4], Vincent Garcia[4], Bruno Bérini[3], Vincent Polewczyk[3], Yves Dumont[3], Sylvia Matzen[1], Philippe Lecoeur[1], Svetlana Santer[2], Thomas Maroutian[1†]*

**Corresponding authors**

* alban.degezelle@universite-paris-saclay.fr

† thomas.maroutian@cnrs.fr

**Author addresses**

[1] Centre de Nanosciences et de Nanotechnologies, CNRS, Université Paris-Saclay, 91120 Palaiseau, France

[2] Institute of Physics and Astronomy, University of Potsdam, 14476 Potsdam, Germany

[3] Université Paris-Saclay, UVSQ, CNRS, GEMaC, 78000 Versailles, France

[4] Laboratoire Albert Fert, CNRS, Thales, Université Paris-Saclay, 91767 Palaiseau, France

## Abstract

Tailoring at will polar textures in ferroelectrics is critical for the development of nanoscale electronics and functional oxide technologies. Freestanding ferroelectric membranes have enabled studies of strain-induced polarization responses, but the control over membrane shape and local polarization typically remains limited to spontaneous buckling or uniaxial mechanical deformations. In this work, we employ a versatile photosensitive-polymer patterning approach to impose programmable bending strain profiles in ferroelectric membranes. Using $BaTiO_3$ as a model system, we demonstrate deterministic 90° polarization rotation driven by engineered in-plane strain, and 180° polarization reversal arising from flexoelectric coupling through a controlled strain gradient. These results establish this programmable bending as a powerful approach to investigate strain-dependent domain structures, leverage flexoelectric effects, and engineer custom ferroelectric landscapes across a wide range of oxide membranes.

## 1. Introduction

Ferroelectric materials have been the subject of extensive research due to their distinctive properties, with a wide range of applications from capacitors to data storage technologies. Beyond its fundamental significance, the ability to pattern ferroelectric domains on-demand opens up transformative opportunities for non-volatile memories[1], flexible electronics[2,3], tunable photonic devices[4], and neuromorphic computing architectures[5]. A prominent strategy for modulating the ferroelectric landscape is strain engineering. This is commonly achieved in thin films by introducing misfit strain through epitaxial growth on a suitable substrate[6,7]. Such an approach enables the tailored adjustment of critical ferroelectric properties, including polarization magnitude, orientation, and switching coercive field.

However, conventional strain engineering in thin films is fundamentally limited by the availability, crystalline structure, and geometry of possible substrates and epitaxial electrodes. Recently, the field has overcome this bottleneck with the emergence of freestanding ferroelectric membranes[8,9]. Through selective etching of a sacrificial layer, the active ferroelectric film can be released from its growth substrate and transferred onto a wide range of host platforms, including bendable ones. This breakthrough has not only accelerated research into the integration of ferroelectric oxides with silicon technology[10,11] but has also unlocked strain engineering possibilities unconstrained by the crystallographic and geometric limitations of traditional oxide substrates.

Numerous studies have investigated the response of freestanding ferroelectric membranes to large bending strains[12–15], exceeding by far those attainable in epitaxial systems. Such extreme morphological deformations are frequently allowed by strain-driven polarization reconfigurations[16], which are highly sensitive to both the nature and intensity of the applied mechanical strain. When shaped into rolls[17,18], wrinkles[19], or folds[14,16,20], these membranes exhibit significant alterations in their ferroelectric landscape. Pioneering experiments have relied on transfer methods that may have compromised spatial precision and reproducibility, but laid critical groundwork for understanding strain-mediated polarization control.

In addition to bending deformations, membranes have also been subjected to homogeneous in-plane tensile strain [21–23] by transferring them onto flexible substrates such as polyimide tapes or plastic films. These methods provide a uniform macroscale strain control, lacking the ability to achieve localized tuning with defined built-in strain. Periodic strain patterns can be generated at the microscale. For instance, this micro-patterning involves a polymer-based technique, such as polydimethylsiloxane (PDMS)[24]. In this approach, a ferroelectric membrane is transferred onto a pre-stretched or heated PDMS substrate. When the applied tensile or thermal stress is released, the relaxation of the PDMS generates a periodic wrinkling pattern in the attached membrane[19,25–27]. Experimental results have shown that this PDMS wrinkling can organize the polarization of ferroelectric $BaTiO_3$ (BTO) membranes into in-plane stripes, producing head-to-head or tail-to-tail domain configurations[28]. Nevertheless, a full 180° polarization reversal in ferroelectric membranes via the flexoelectric effect remains extremely challenging. This typically requires either huge bending deformations[14] or the application of a strong, localized tip pressure to generate the large strain gradients required[25,29–32]. These methods remain limited in both geometric flexibility and tunability. They lack the ability to selectively target specific regions within the membrane, and the resulting wrinkle patterns are inherently dictated by the direction and extent of the initial pre-stretch without any post tunability. Advanced spatially selective strategies for ferroelectric strain engineering would be capable of delivering continuously tunable, as well as local microscopic control over strain, polarization orientation, and domain structure without compromising the material's structural and functional integrity.

In this study, we introduce a robust and highly tunable platform for strain-induced polarization patterning of ferroelectric membranes. At the core of our method is an optically defined deformation process, where controlled membrane bending is achieved using a photosensitive polymer layer dynamically deformed when illuminated by a spatial light modulator (SLM). This SLM-based optical control allows for programmable, site-specific shaping with micrometer-scale precision. Thanks to its tunability, we demonstrate deterministic 90° and 180° polarization rotations in an archetypal BTO membrane, through in-plane strain and flexoelectric coupling, respectively. In order to showcase this approach's versatility, we extend its applicability to other oxide membranes, such as dielectric $SrTiO_3$ (STO) and ferroelectric $Pb(Zr,Ti)O_3$ (PZT). The resulting polarization configurations and domain patterns are mapped using piezoresponse force microscopy (PFM), offering direct insights into the polarization landscapes and strain-mediated switching mechanisms at play. By overcoming the geometric rigidity of previous approaches, our platform enables an unprecedented freedom in engineering local strain, domain structures, and polarization orientations, opening new avenues for the design of reconfigurable ferroelectric devices.

## 2. Results

### 2.1. On-demand patterning of oxide membranes

To achieve on-demand patterning of oxide membranes, we leverage the unique advantages of polymers, including mechanical flexibility, chemical resistance, optical transparency, and dielectric behavior. Photosensitive polymers[33] in particular, offer an additional benefit, because optical illumination enables precise spatial remote control, scalability, and tunability, making them ideal for patterning processes[34]. One such polymer is the poly {1-[4-(3-carboxy-4-hydroxyphenylazo) benzenesulfonamido]-1,2-ethanediyl, sodium salt} (PAZO), comprising azobenzene chromophores covalently tethered to an optically inert backbone. PAZO is renowned for its vectorial photomechanical response[35]. Under polarized-light excitation, azobenzene side chains undergo reversible trans–cis photoisomerization, driving their alignment perpendicular to the incident electric-field vector. This local ordering of chromophores produces anisotropic stresses within the polymer matrix. The accumulation of these stresses results in a macroscopic surface deformation, with magnitude determined by light intensity, light polarization, and matrix viscoelasticity. To harness this mechanism for on-demand, spatially resolved surface patterning, we utilized a spatial light modulator (SLM). This SLM projects a custom light polarization pattern onto the sample, which the PAZO film then converts into a corresponding topographic deformation[36].

In the present work, we employ this optical platform to imprint deformations into crystalline oxide membranes transferred on PAZO, exploiting their anisotropic crystallography to generate highly controlled and high-magnitude strain patterns. The oxide membrane conforms precisely to the polymer's surface relief, and our approach provides significantly greater versatility compared to conventional substrate-clamped strain engineering. We first introduce our fully tunable, high-precision photomechanical apparatus and then demonstrate its capacity to spatially pattern the ferroelectric polarization landscape through bending strain in prototypical ferroic oxides.

Oxide membranes were fabricated from the archetypal “incipient’ ferroelectric STO, the archetypal lead free ferroelectric BTO, and the lead-based ferroelectric PZT. Each membrane included an epitaxial $SrRuO_3$ (SRO) bottom electrode, deposited in situ during thin-film growth. The fabrication employed a sacrificial-layer scheme, with acid-etched $La_{2/3}Sr_{1/3}MnO_3$ (LSMO)[37] for BTO and PZT, and water-soluble $SrVO_3$ for STO[38]. After growth, the oxide layer and its bottom electrode were released from the original substrate and transferred with the help of a PDMS stamp onto a glass support coated with a PAZO thin film. Details of thin film growth, membrane release, and polymer preparation are provided in the **Methods** section. The overall growth-transfer process is depicted in **Figure 1a**. Upon completion of the transfer, the oxide membrane of interest rests on top of the PAZO layer, with its epitaxially attached bottom electrode in direct contact with the polymer.

The patterning of oxide membranes transferred atop of PAZO is then realized with the SLM-based digital polarization setup (**Figure 1b**). A 491 nm continuous wave laser is expanded to fully illuminate the active area of the SLM. According to the grayscale map addressed to the SLM, a complex polarization pattern is created. This polarization pattern is then demagnified and projected onto the oxide/polymer/glass sample, inducing a deformation of the azo-polymer, ultimately deforming the oxide on top of it. Exposure is performed from the glass side of the sample. **Figure 1c** shows a panel of realized patterns in the oxide membrane topography, displayed in 3D for clarity. A more detailed description of the optical setup and surface relief gratings formation is provided elsewhere[36,39,40].

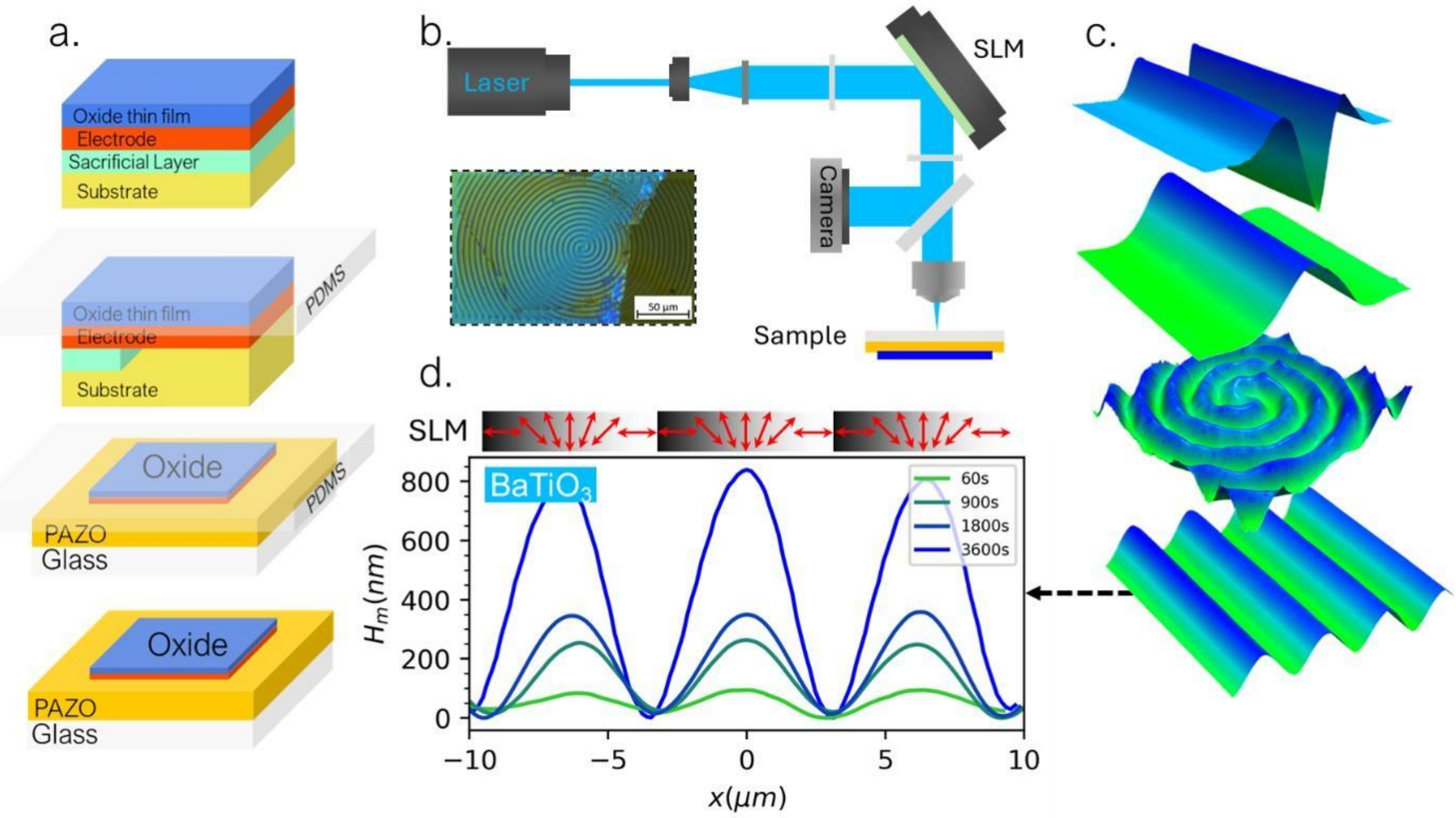

**Figure** 1: (a) Sketch of the fabrication process of the freestanding membrane on the PAZO photosensitive polymer (from top to bottom): Stack growth, membrane release, transfer to PAZO and removal of the PDMS stamp. (b) Optical exposure setup, the inset shows an exposure spot on the membrane edge, with deformation patterns imprinted in the polymer alone (right) and in the polymer + membrane stack (left). (c) 3D representation of the oxide membrane topography after patterning (from top to bottom): Deep valley, giant hill, radial grating vortex, and 1D hill & valley grating. (d) Evolution of a BTO membrane deflection for different exposure times. The top inset shows the grayscale pattern input into the SLM, the corresponding light polarization orientation with respect to the grey scale value is depicted by the red arrows.

This method provides an unprecedented pattern control through adjustable parameters: amplitude, location, pattern shape, and period. To showcase its capabilities, we imprinted a periodic pattern of hills and valleys (H&V) with a 6 μm period onto a BTO membrane. As illustrated in **Figure 1d**, the membrane topography can be dynamically tuned by varying the exposure time, from flat to an amplitude of up to 880 nm. For these patterning conditions, one hour of exposure is sufficient to reach the saturation stage in polymer reconfiguration[41]. A typical exposure process was recorded and presented in an accelerated format in **Supplementary Video 1**.

The pattern periodicity and shape can be adjusted by applying distinct grayscale control patterns to the SLM. For the H&V configuration, the corresponding grayscale pattern is displayed above the graph in **Figure 1d**, with red arrows indicating the orientation of the incident light polarization in the substrate plane. This approach allows for localized patterning simply by directing the exposure spot to the target area. Furthermore, the feature size can be simply tuned by modifying the focusing optics.

To maximize the induced strain of the supported oxide membrane, it is desirable to achieve the largest PAZO deflection over the smallest pattern period. This geometry minimizes the H&V curvature radius and consequently maximizes the bending strain in the membrane. However, the polymer maximum deflection is constrained by factors such as pattern shape, period, and exposure time. For the H&V pattern shown in **Figure 1d**, the optimal period for achieving maximum deflection in the bare polymer (without a membrane) lies between 5 and 6 μm, as measured by atomic force microscopy (AFM) (**Figure S1**) and in agreement with previous observations[41]. This intrinsic constraint limits the maximum strain achievable and may be further influenced by additional factors related to the added membrane (alignment with the membrane crystallographic axes, membrane thickness, oxide mechanical properties…). The method offers exceptional versatility, enabling on-demand shaping of a wide variety of materials that can be fabricated into membranes. For instance, **Figure 2a** shows the realization of H&V patterns (6 μm period, 1 hour exposure) in thin membranes of STO and PZT, in addition to BTO (details on membrane and electrode thickness are provided in the **Methods** section). Remarkably, we achieve deflection heights of up to approximately 1 μm for this specific pattern across all membranes, underscoring the technique's robustness and adaptability. These large deflections make it possible to create tunable highly strained periodic patterns. To estimate the achieved in-plane strain ($\varepsilon_{xx}$), we applied the Euler-Bernoulli beam theory (see **Supplementary Information**) to compute a one-dimensional (1D) strain profile on the membrane topography, as shown in **Figure 2b**. With this H&V pattern, we reached bending strains up to about 1% in both tensile (red) and compressive (blue) regimes, periodically distributed across the pattern.

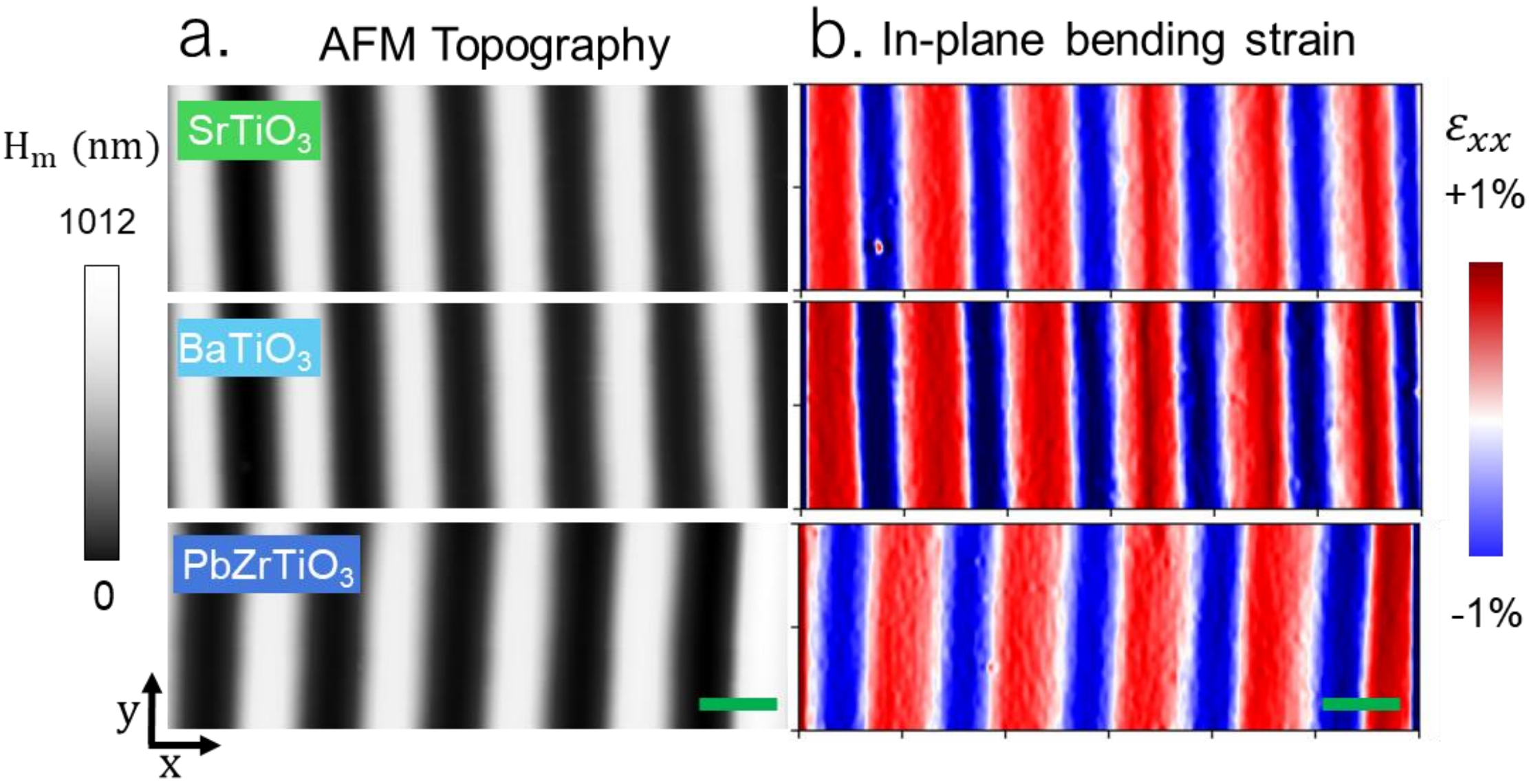

**Figure** 2: (a) AFM topography of STO, BTO and PZT membranes patterned with a hill & valley grating shape. (b) The associated strain map of the in-plane bending strain $\varepsilon_{xx}$ computed from the AFM topography. The topography induces an alternating periodic array of compressive (blue) and tensile (red) regions imposed by bending. Green scale bar is 6 μm long.

### 2.2. Periodic patterning of the polarization on a ferroelectric BTO membrane

Both theoretical predictions and experimental studies suggest that the strain levels achieved through our patterning of ferroelectric materials such as PZT or BTO (**Figure 2**) should impact polarization landscapes[42,43]. Dealing with the STO case, the induced strain remains insufficient to trigger ferroelectricity at room temperature[23,44], as confirmed by the lack of any measurable piezoresponse (**Figure S2)**.

To investigate the polar landscape of the H&V-patterned BTO membrane, we need both vertical and lateral PFM (VPFM and LPFM) mapping. In its as-grown state, the BTO layer displays a uniform, upward polarization, defined as +c-oriented along the vertical crystallographic c-axis, typical of its tetragonal phase. This was confirmed by X-ray diffraction (XRD, **Figure S3**), which revealed the (002) reflection at 44.75° (c = 0.4045 nm), slightly exceeding the bulk value of 0.4036 nm[45]. After transfer, the BTO membrane retains its +c polarization (**Figure 3a**), as further verified by PFM measurements on the flat membrane (**Figure S4**). This out-of-plane upward polarization might be stabilized by the presence of the epitaxial SRO bottom electrode.

After patterning, VPFM and LPFM images of the wrinkled BTO membrane are shown in **Figure 3b**, along with corresponding topography. The PFM images display only three periods, but this patterned behavior extends over more than 10 periods, with stripes spanning nearly the entire SLM window (about 80 µm in length in this specific pattern). The PFM cantilever was oriented parallel to the wrinkles in a so-called 0° configuration, making LPFM primarily sensitive to IP $a_1$-axis oriented domains, aligned along the applied bending strain and orthogonal to the cantilever axis.

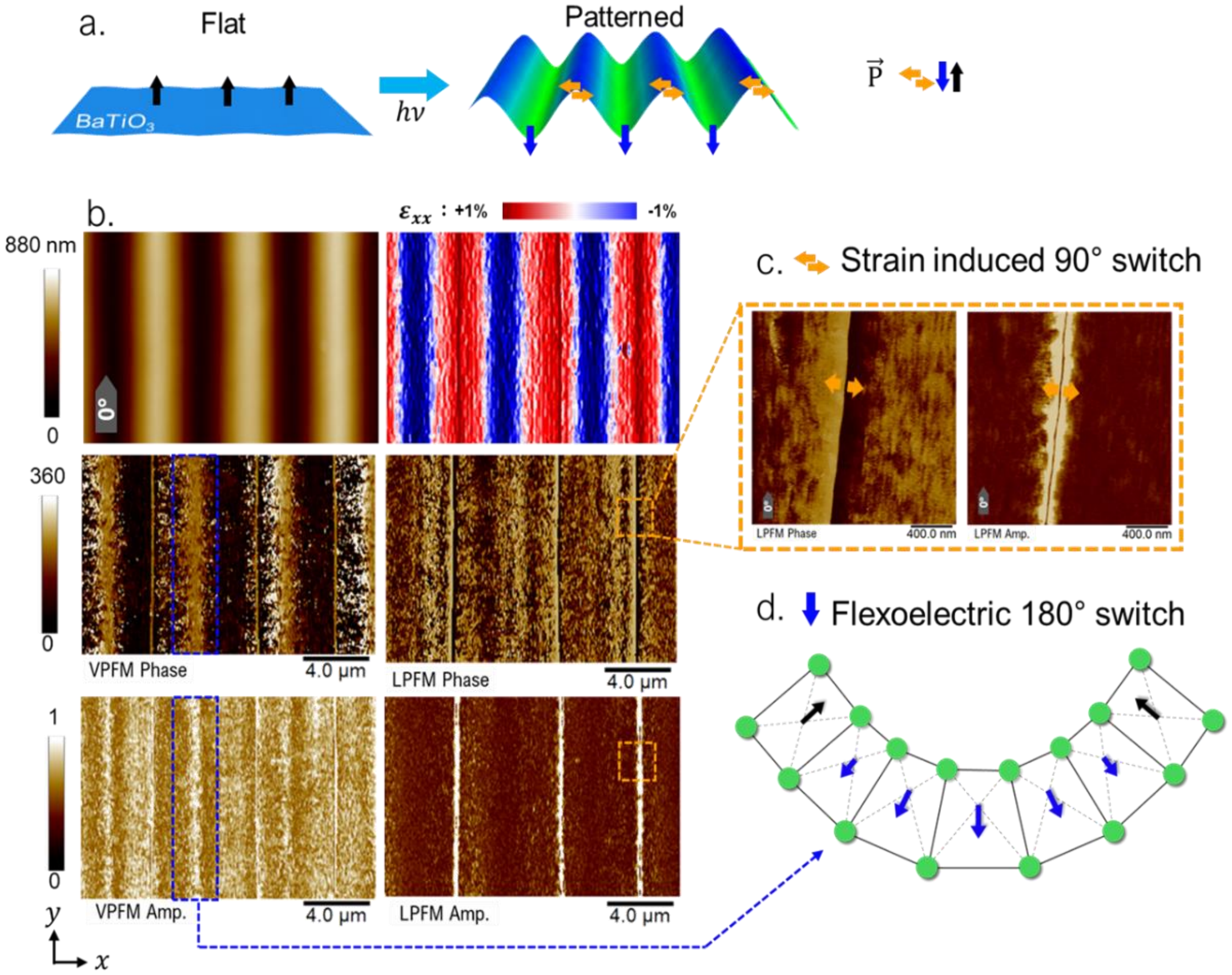

**Figure** 3: (a) Illustration of the BTO membrane polarization and shape, before and after patterning. The transferred BTO membrane is single domain (+c, black arrow) before patterning. (b) Topography, VPFM and LPFM images done after patterning on the BTO membrane H&V pattern with the associated in-plane strain map with tensile (compressive) region in red (blue). (c) LPFM zoom on top of a BTO hill: the strong IP strain triggered a 90° polarization switch, the polarization is now IP organized in two tail-to-tail domains, separated by a clear domain wall with no LPFM amplitude. (d) Illustration of the flexoelectric switch in the BTO: the strong downward flexoelectric polarization (blue arrow) overcomes the spontaneous polarization (black arrow) progressively in the valleys triggering a 180° phase reversal in the VPFM channel, accompanied by a strong amplitude. Unstrained areas on the sides of the hill exhibit the spontaneous upward polarization of BTO grown on an SRO electrode.

At the wrinkle crests, we observe striking LPFM amplitude stripes, which correlate with a distinctive domain structure in the LPFM phase image. The zoomed inset in **Figure 3c** reveals a 90° polarization rotation, resulting in an in-plane orientation of the ferroelectric polarization on top of the BTO hill. This rotation generates two large tail-to-tail (T-T) domains, separated by a sharp domain wall that is clearly visible as a drop in LPFM amplitude between the domains. Averaged cross-sectional scan profiles, presented in **Figure 4b** for both topography and LPFM signals, confirm these observations. On the hillside, where the polarization retains its spontaneous +c orientation (black arrow), no LPFM amplitude is detected. Conversely, at the hilltop, subject to a maximum IP tensile strain of ~0.7%, a pronounced LPFM amplitude peak emerges. This peak is bisected by a distinct domain wall, where the amplitude drops to zero. Across this domain wall, the LPFM phase undergoes a 180° rotation, as evidenced by the clear phase shift at the wall location. The orientation relative to the cantilever torsion is illustrated by the orange arrows. These features collectively represent a clear signature of a 90° polarization switch at the hill apex, characterized by a distinctive T-T striped domain configuration.

In addition to the 90° polarization rotation on top of the hills, we observe an additional phenomenon occurring, this time, in the valleys. In the VPFM scans (**Figure 3b**), within the valleys, we detect pronounced VPFM amplitude stripes correlated with a 180° phase reversal, transitioning from dark (+c, upward polarization) to bright (-c, downward polarization). The averaged cross-sectional profiles of the VPFM scans (**Figure 4a**) offer further insight into the reversal behavior. Along the wrinkle slopes, where IP strain is reduced, the VPFM amplitude reveals a measurable piezoresponse. As the PFM tip approaches the valley, the response initially diminishes but, near the deepest point, exhibits a sharp increase, even surpassing the amplitude observed on the slopes. Additionally, the VPFM phase (**Figure 4a**, middle panel) demonstrates a gradual switching from upward to downward polarization. At the midpoint of this reversal (corresponding to a 90° rotation), the VPFM amplitude drops to zero. This 180° phase reversal, accompanied by a localized amplitude spike, is consistently observed across all the examined valleys.

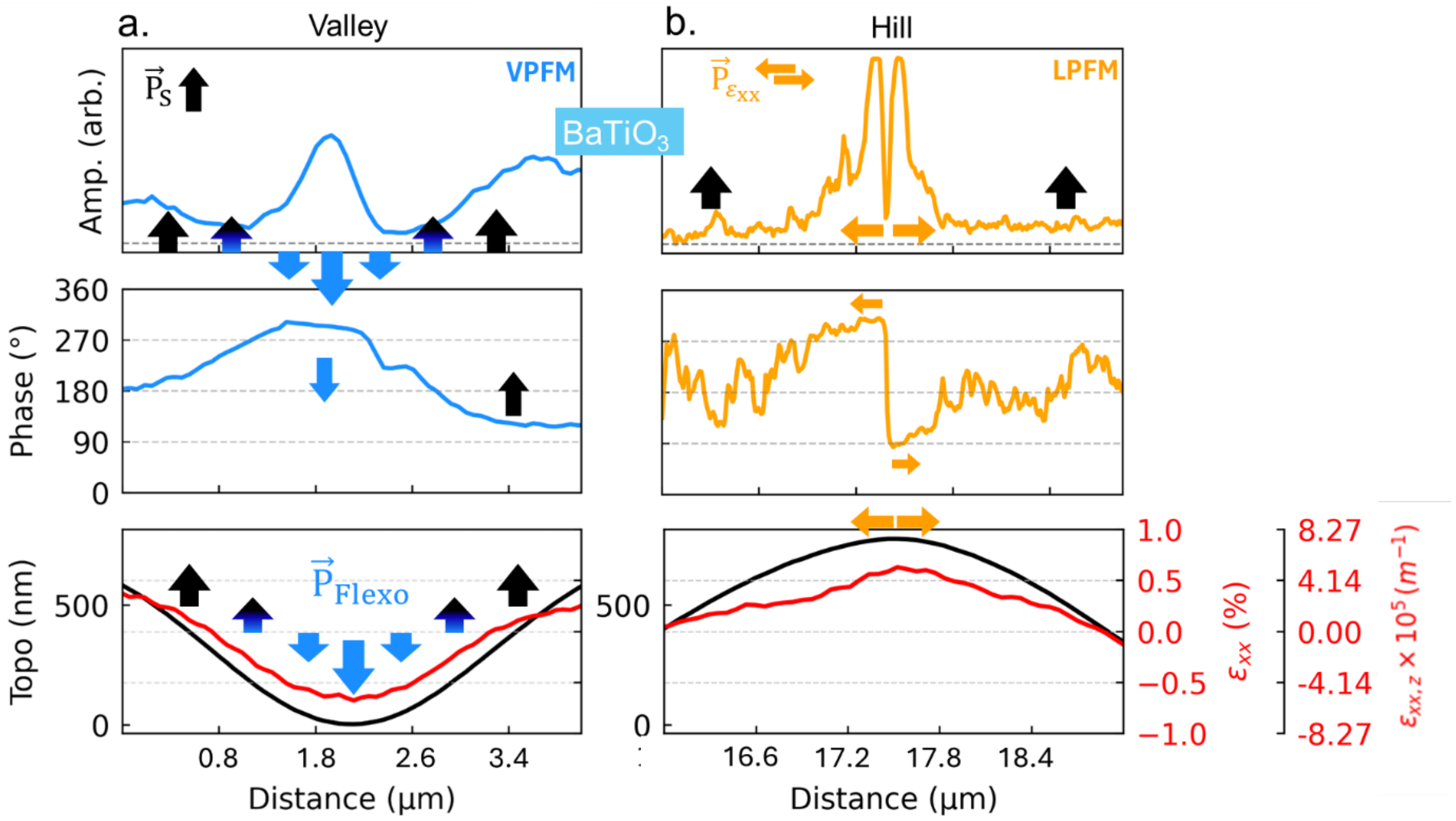

**Figure** 4: (a) VPFM signal (blue line) in a BTO valley with amplitude (top panel) and phase (middle panel). Arrows depict the flexoelectric polarization component ($P_{Flexo}$, in blue) and the spontaneous upward one ($P_S$, in black). The AFM height profile is shown in the bottom panel (black line), together with the computed IP bending strain (red line). Note that the IP strain gradient has the exact same curve, albeit with a separate scale as indicated on the bottom right. (b) LPFM signal (orange line) on a BTO hill with amplitude (top panel) and phase (middle panel), showing a clear 180° phase contrast separated by a domain wall with no LPFM amplitude. The IP polarized region ($P\varepsilon_{xx}$, orange arrows) is very narrow (~400 nm width). Outside this area, the polarization is out-of-plane upward-oriented (black arrows). The critical strain triggering the 90° polarization switch is found around +0.45%.

## 3. Discussion

### 3.1. Flexoelectricity-driven 180° polarization reversal

Based on the piezoelectric effect, the compressive strain experienced in the valleys by the BTO membrane should enhance its pristine +c polarization. However, this expectation does not match our experimental observations, which consistently indicate a flexoelectricity-induced polarization reconfiguration. For the 800 nm-deflection HV patterns, we computed the IP strain gradient along the membrane thickness, $\varepsilon_{xx,z}$. These calculations were performed using the mechanical model detailed in the **Supplementary Information**, applied to the patterned BTO topography.

By averaging the data across the entire PFM image along the wrinkle length, we find that the valleys exhibit an average maximum gradient of approximately $-6.5 \times 10^5$ $m^{-1}$, as shown in **Figure 4a** (bottom panel), and can locally reach $-1 \times 10^6$ $m^{-1}$. The evolution of the VPFM amplitude across the valleys (**Figure 4a**, top panel) shows an initial decrease followed by a strong spike at the bottom of the valley. The spontaneous upward polarization is opposed to the flexoelectric polarization induced in the valleys, which aligns downward due to a negative strain gradient. Initially, the VPFM amplitude diminishes progressively as the strain gradient intensifies the flexoelectric polarization component. Consequently, the total polarization gradually deviates from the +c-axis orientation. Eventually, flexoelectric polarization overcomes spontaneous polarization, resulting in a complete 180° phase reversal. This is depicted in **Figure 4a** by the colored arrows representing spontaneous (in black) and flexoelectric (in blue) contributions to the polarization. We also observed that a lower deflection of the H&V pattern gives a lower flexoelectric effect, as shown in the PFM data of **Figure S5** and **Figure S6**.

These observations further corroborate that the strain gradient serves as the primary mechanism driving polarization reversal within the BTO valleys. By precisely tuning the polymer morphology, and consequently the membrane geometry, the magnitude of the strain gradient can be modulated to control polarization orientation.

Flexoelectric 180° reversal has been extensively studied in clamped thin films and two-dimensional materials, and more recently it has also been investigated in freestanding membranes. In the latter case, membrane bending enables the generation of strain gradients with magnitudes comparable to those produced by tip forces or strain relaxation processes.

Indeed, the polarization rotation observed in our patterned BTO membrane aligns with the findings of *Dong et al.*[13], who reported a similar continuous polarization rotation in rolled BTO membranes. Such rotation has been attributed to the flexoelectric effect induced by bending.

Other previous studies have demonstrated that large strain gradients ($\sim 10^7$ $m^{-1}$) can lead to giant flexoelectric enhancements of the polarization in thin $BiFeO_3$ (BFO) and STO folds[14]. In PZT thin films, tip-induced pressure has been shown to generate strong enough gradients to trigger a flexoelectric polarization reversal directly observed by PFM[31]. When the gradient at stake is in the $10^4$ – $10^5$ $m^{-1}$ range, indirect effects of flexoelectricity have been reported to influence the properties of ferroelectric membranes. For instance, when the flexoelectric polarization is opposed to the film spontaneous polarization, modifications in photoconductance were observed in BFO[46], while strain-mediated polarization rotation was triggered at low applied strain in PZT[16]. Furthermore, periodic wrinkles in BTO have shown changes in coercive voltage and imprint bias[19,26]. These studies highlight changes in the material properties and were correlated with the curvature, albeit without triggering flexoelectric 180° polarization switching. Freestanding ferroelectrics have also been reported to exhibit reduced switching voltages[47], and 180° polarization reversal may occur at slightly lower strain gradients than in clamped films[29].

Strain gradient amplitude is not the sole determining factor; the transverse flexoelectric coefficient $\mu_{12}$ also plays a crucial role. However, obtaining precise values for this coefficient remains challenging, as they can vary depending on the measurement method. For instance, BTO is generally believed to exhibit a stronger flexoelectric coefficient compared to PZT or STO [48,49], potentially requiring lower strain gradients to trigger flexoelectric switching. We transferred a thin PZT membrane (14 nm thickness) onto a PAZO substrate. In its flat state, this membrane was fully polarized downward in this case (–c) as shown in **Figure S4**. We then patterned it using the same high deflection H&V pattern and performed PFM mapping, as for BTO. The corresponding topography, strain map, and PFM scans are shown in **Figure S7**. Despite having a similar deflection height, and therefore a comparable strain gradient to that obtained in BTO, we did not observe any change in the VPFM response. In particular, we expected a flexoelectric-induced polarization reversal on top of the hills this time, since the spontaneous downward polarization is opposed to the upward flexoelectric field in these regions. This absence of switching suggests that thin BTO membranes exhibit a superior flexoelectric coefficient compared to thin PZT. *Cross et al.* reported that the flexoelectric coefficient of PZT is around 1.4 μC/m[50], while it reaches several μC/m for BTO[51]. Recent studies corroborate these trends, showing that this coefficient in ultrathin PZT membranes may be even lower[52]. Another plausible explanation for the lack of 180° switching in PZT could be the competition between the in-plane strain, which tends to favor in-plane polarization, and the flexoelectric effect, which tends to reverse the out-of-plane spontaneous polarization. While the reduced thickness of the PZT membrane was chosen deliberately to avoid any 90° in-plane switching on the hills, there still is a maximum in-plane strain of ~0.25% at the hill apex (**Figure S7b**). This interplay likely results in no net polarization switching for this geometry. In contrast, in the case of BTO, compressive strain in the valleys did not prevent the downward polarization reversal. Furthermore, we note that the STO membrane patterned the same way (**Figure 2**) did not show any sign of flexoelectric induced effect, as probed by PFM (**Figure S2**). This suggests that STO also has a lower flexoelectric coefficient than BTO.

### 3.2. Strain-induced 90° polarization rotation

Analogous structures such as the in-plane stripe from **Figure 3c** have recently been reported in BTO with wrinkled morphologies [20,28]. Strain-induced 90° polarization rotation has also been observed in bent configurations such as PZT and BTO folds and in PTO wrinkles structures [16,20,53]. Such a 90° switch on top of the BTO hills leads to a supposedly charged domain wall, here between two T-T domains. Despite the clear presence of an in-plane domain wall, evidenced by the LPFM amplitude drop at the LPFM phase shift (**Figure 4b**), the VPFM scan of this region (**Figure S8)** reveals a small and sharp amplitude stripe precisely at the wall location. This stripe is accompanied by a bright phase contrast, suggesting a potential downward polarization. As shown in the cross-sectional plot of the PFM scans (**Figure S8b**), the VPFM amplitude spike is confined to a narrow region at the domain wall, where the LPFM amplitude drops to zero. We rule out any crosstalk between the VPFM and LPFM channels due to tip misalignment, as the wall is located exactly on the crest with no net tilt observed, as confirmed by the zoomed profile in **Figure 4b**.

This signature could resemble a non-Ising-like[54] domain wall between two T-T IP-polarized domains. However, this scenario does not fully align with our observation. It is arguable that the upward flexoelectric field on top of the hill could, in principle, induce this type of non-Ising behavior[55]. But rather than rotating downward (as observed on the bright VPFM phase stripe in **Figure S8a**), the positive gradient would induce an Ising-Néel like rotation in the same upward direction as the flexoelectric polarization. An alternative explanation could involve the presence of a ferroelectric monoclinic (m1) phase at the hill apex, where the polarization components lie along both the c- and a-axes[6]. The critical strain required to induce the observed polarization switch on the BTO hill is ∼ +0.5% (**Figure 4**). According to thermodynamic studies, this level of strain, under biaxial misfit conditions, would typically stabilize a purely in-plane polarization[56]. In our case, however, the uniaxial tensile strain, combined with the upward flexoelectric component, may drive a deviation from purely IP polarization toward a monoclinic m-phase[57]. While this peculiar domain wall region warrants further local-scale analysis, for instance with atomically resolved Scanning Transmission Electron Microscopy, such an investigation lies beyond the scope of this paper.

Our periodically patterned BTO serves as an ideal model system to reveal the complex interplay between strain and strain gradient, exhibiting two distinct polarization regimes governed by different driving mechanisms. This is emphasized in **Figure 5**, where the evolution of the PFM amplitude is plotted as a function of the in-plane strain and strain gradient. At strain levels below ~0.45%, the strain gradient dictates the polarization state. In this regime, compressive regions promote a reorientation of the upward polarization into the downward direction, as reflected by the initial suppression and subsequent sharp increase of the VPFM amplitude past a negative gradient threshold of about $-5 \times 10^5$ $m^{-1}$. This behavior is reminiscent of a result from *Lu et al.*[30] where they used AFM tip force to trigger the flexoelectric reversal. Conversely, in the low-tensile regime, the strain gradient reinforces the upward polarization, leading to an enhancement of the VPFM amplitude between 0 and +0.25% strain. Once the strain exceeds ~0.45% tensile, the strain itself becomes the dominant control parameter despite large upward flexoelectric polarization. This crossover is manifested by an abrupt polarization rotation, evidenced by the sharp rise in LPFM amplitude accompanied by a reduced VPFM response. In the narrow strain range between 0.25% and 0.45%, the flexoelectric and strain contributions may nearly cancel each other, giving rise to this transition region. Together, these results demonstrate a strain-gradient to strain-driven transition of polarization control in BTO thin films membranes.

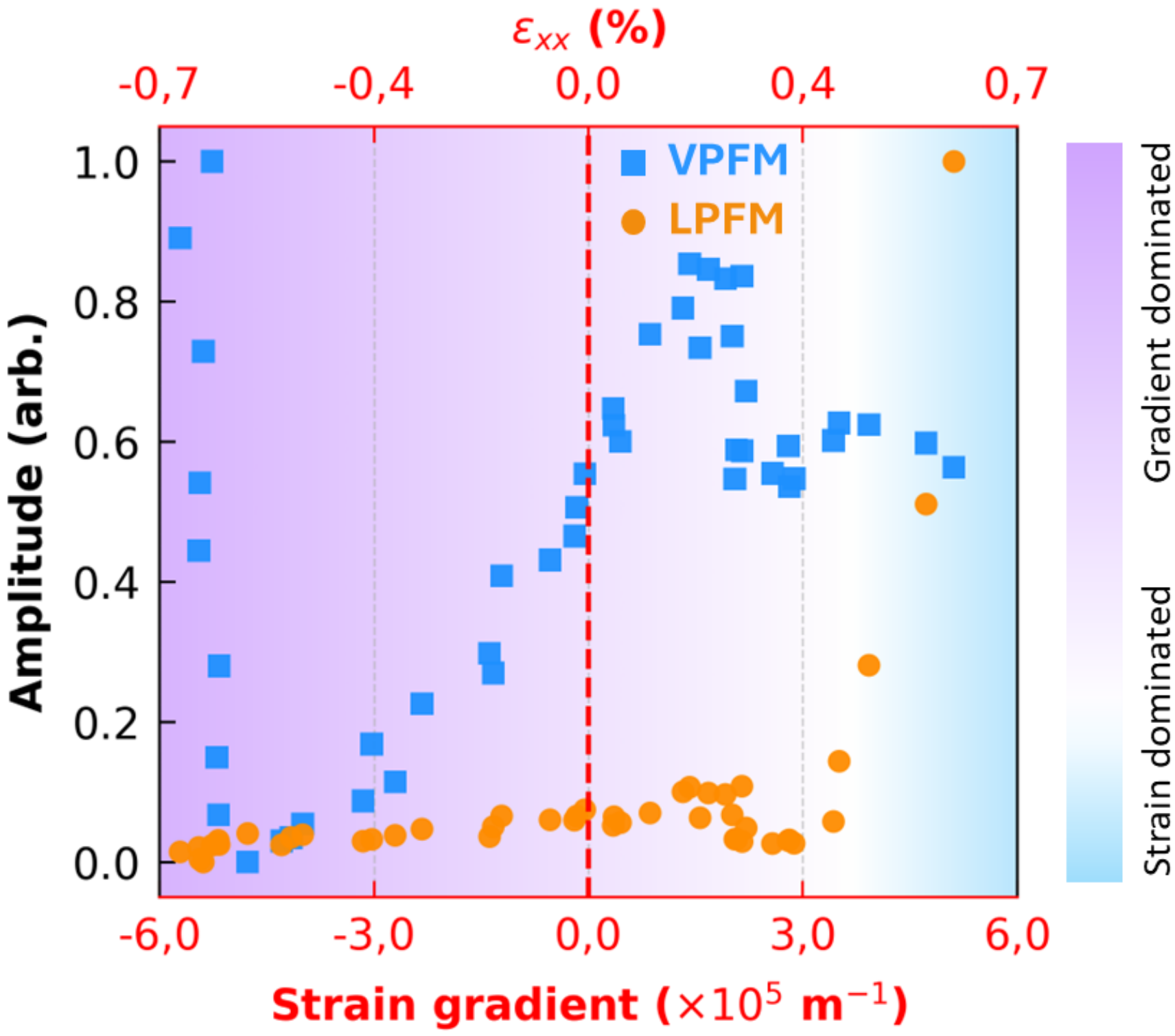

**Figure 5**: Evolution of the VPFM (blue) and LPFM (orange) amplitudes in BTO membrane as a function of in-plane strain (top axis) and strain gradient (bottom axis). The evolution of both respective amplitudes highlights a gradient dominated regime (purple background) below +0.25% strain, and a strain-dominated regime (light blue background) above +0.45%. The displayed strain/gradient range was calculated going from a valley depth to a hill top on an AFM height profile.

## 4. Conclusion

Our results show that optically driven membrane deformation provides a powerful route to achieve on-demand mechanical control of polarization orientation in freestanding oxide membranes, enabling programmable topographies with micrometer-scale precision and generating both substantial in-plane strains and well-defined strain gradient patterns. Leveraging these strain fields, we achieved precise control over polarization reorientation, including 90° in-plane rotation and, in BTO, a complete 180° flexoelectric-driven polarization reversal. Our findings highlight that the stability and switching dynamics of these engineered domain structures are highly sensitive to the interplay between local strain, strain gradient, and material-specific flexoelectric coefficients, as evidenced by the contrasting behaviors observed in BTO, PZT and STO. Beyond advancing fundamental studies of flexoelectricity and strain-polarization coupling, this strategy establishes a platform for reconfigurable ferroelectric architectures, free from the constraints of mechanical clamping. This capability unlocks new possibilities, from designing flexible nanoelectronics to generating complex topological domain structures, where polarization landscapes can be dynamically written, and potentially erased, with optical precision.

## Methods

### Thin film growth

The growth of functional oxides with electrodes, on substrate buffered by a sacrificial layer was performed by pulsed laser deposition using a 248 nm KrF laser.

**$GdScO_3$(011)/$La_{2/3}Sr_{1/3}MnO_3$(20nm)/$SrRuO_3$(10nm)/$BaTiO_3$(20nm)**: Temperature of the substrate was 650°C, under a dynamic pressure of 0.15 mbar with $O_2$ gas. The laser repetition rate was 5 Hz for SRO, PZT, 2 Hz for LSMO and 4 Hz for BTO, at a common laser fluence of 2.5 J/cm$^2$ and a target-substrate distance of 5 cm. After growth, the sample was cooled down to room temperature under 80 mbar of $O_2$.

**$SrTiO_3$(001)/$La_{2/3}Sr_{1/3}MnO_3$(20nm)/$SrRuO_3$(8nm)/$Pb(Zr_{0.2},Ti_{0.8})O_3$(22nm) (Upward)**: Growth performed at a substrate temperature of 630°C (SRO and LSMO) and 600°C for PZT, under a dynamic pressure of 0.15 mbar with $O_2$ gas for SRO and LSMO, and with $N_2O$ gas for PZT at 0.15 mbar. The laser repetition rate was 5 Hz for SRO and PZT, 2 Hz for LSMO, at a common laser fluence of 2.5 J/cm$^2$ and a target-substrate distance of 5 cm. After growth, the sample was cooled down to room temperature under 300 mbar of $O_2$.

**$SrTiO_3$(001)/$La_{2/3}Sr_{1/3}MnO_3$(20nm)/$SrRuO_3$(8nm)/$Pb(Zr_{0.2},Ti_{0.8})O_3$ (14nm) (Downward)**: The growth was performed at a substrate temperature of 630°C (SRO and LSMO) and 650°C for PZT, under a dynamic pressure of 0.15 mbar with $O_2$ gas for SRO and LSMO, and with $O_2$ gas for PZT at 0.2 mbar. The laser repetition rate was 5 Hz for SRO and PZT, 2 Hz for LSMO, at a common laser fluence of 2.5 J/cm$^2$ and a target-substrate distance of 5 cm. After growth, the sample was cooled down to room temperature under 0.2 mbar of $O_2$.

**$SrTiO_3$(001)/$SrVO_3$(8nm)/$SrTiO_3$(3nm)/$La_{2/3}Sr_{1/3}MnO_3$(5nm)/$SrTiO_3$(15nm)**: Substrate temperature was 760°C for all the layers, under a dynamic pressure of $O_2$ gas, with a chamber pressure of 5 x $10^{-6}$ mbar for SVO and LSMO, and 5 x $10^{-6}$ mbar for STO layers, respectively. The laser repetition rate was 2 Hz for all the growths, at a common laser fluence of 1.85 J/cm$^2$ and a target-substrate distance of 5 cm.

**Membrane fabrication process**

Freestanding membranes of various oxide materials were released via the sacrificial layer route. For samples of PZT and BTO we employed LSMO as a sacrificial layer. The as-grown samples were attached to a ~300 µm-thick PDMS 10:1 (in mass) stamp and then immersed in an etching solution of 200 mL of deionized water with 5 mL of HCl and 50 mg of KI. After two days, the LSMO is completely etched and the film is released from the growth substrate, attached to the PDMS stamp.

For the STO sample where the sacrificial layer was SVO, the same PDMS 10:1 stamp was used. After stamping the sample with the PDMS, the sample was immersed in 45°C DI water. After a week, the sample was released from the growth substrate, attached to the PDMS stamp.

In both cases, the PDMS is then stamped onto the PAZO polymeric layer spun on glass. After a gentle pressing, the PDMS stamp is peeled off easily with the full membrane transferred onto the PAZO. The transfer of the oxide membrane onto the PAZO is done immediately after being scooped off the etchant solution and dried with nitrogen gun.

**PAZO Thin film fabrication**

The azobenzene-containing photosensitive polymer PAZO (poly{1-[4-(3-carboxy-4-hydroxyphenylazo)benzenesulfonamido]-1,2-ethanediyl, sodium salt}) is purchased from Sigma-Aldrich. The polymer solution is prepared by dissolving PAZO in a 95/5 volume ratio of methoxy ethanol/ethylene glycol solution. The concentration used is about 200 g/l. The PAZO is then spin coated onto fused silica glass (Neyco) 200 $\mu m$ thick substrates at 8000 rpm for 20 sec to prepare 1.5 $\mu m$ thick PAZO films on glass substrate. The thickness of the polymer was inspected using a Bruker Dektak mechanical profilometer.

**Piezoresponse Force Microscopy**

PFM mappings were acquired with an atomic force microscope (Nanoscope V multimode, Bruker) coupled with two external lock-in amplifiers (SR830, Stanford Research) to record simultaneously the vertical and lateral responses. An AC voltage oscillating at a frequency of 35 kHz was applied between the bottom electrode and a grounded Pt-coated tip. We used cantilevers with stiffnesses ranging from 7 to 40 $N.m^{-1}$. PFM profile data were extracted and analyzed with WSxM[58].

**Polymer patterning with SLM**

Surface patterning was performed with an optical setup based on a spatial light modulator (SLM) acting as a polarization rotator, described in detail elsewhere [36]. Briefly, a reflective liquid-crystal-on-silicon SLM (PLUTO-2.1 from Holoeye, 1920 × 1080 pixels) was positioned between two orthogonal quarter-wave plates, enabling full 0–360° polarization rotation with pixel-level precision. A 491 nm laser beam (Cobolt Calypso) was expanded to uniformly illuminate the SLM (~2 $cm^2$) and then focused onto the polymer film through a microscope objective (40x), producing a spot of ~230 × 130 μm. A dichroic mirror and off-axis camera were used for sample alignment and in situ monitoring of surface relief grating formation.

The surface relief grating within PAZO films remains stable under ambient conditions and is not thermally erasable, although it can be reversed under specific irradiation conditions[57].

**Associated content**

The supporting information is available online. It contains details on the mechanical modelling of bending strain, X-ray diffraction and PFM data on the as-grown samples, as well as additional PFM scans of STO, PZT and BTO wrinkled membranes.

**Acknowledgements**

This work was supported by the French Agence Nationale de la Recherche (ANR) through the FLEXO project (ANR-21-CE09-0046), and by the French national network RENATECH for nanofabrication. It also received funding from the European Union's Horizon 2020 research and innovation program under grant agreement No 964931 (TSAR project). SS, JS and SL gratefully acknowledge financial support from the European Regional Development Fund (EFRE) in the framework of the StaF project *EcoDipol* (ILB Brandenburg, Germany). We thank Sofia Assam for the speed-up video.

## References

(1) Sun, H.; Wang, J.; Wang, Y.; Guo, C.; Gu, J.; Mao, W.; Yang, J.; Liu, Y.; Zhang, T.; Gao, T.; Fu, H.; Zhang, T.; Hao, Y.; Gu, Z.; Wang, P.; Huang, H.; Nie, Y. Nonvolatile Ferroelectric Domain Wall Memory Integrated on Silicon. *Nat. Commun.* **2022**, *13* (1), 4332. https://doi.org/10.1038/s41467-022-31763-w.

(2) Cheng, Y.; Dong, G.; Li, Y.; Yang, G.; Zhang, B.; Guan, M.; Zhou, Z.; Liu, M. Strain Modulation of Perpendicular Magnetic Anisotropy in Wrinkle-Patterned $(Co/Pt)_5/BaTiO_3$ Magnetoelectric Heterostructures. *ACS Nano* **2022**, *16* (7), 11291–11299. https://doi.org/10.1021/acsnano.2c04754.

(3) Bakaul, S. R.; Serrao, C. R.; Lee, O.; Lu, Z.; Yadav, A.; Carraro, C.; Maboudian, R.; Ramesh, R.; Salahuddin, S. High Speed Epitaxial Perovskite Memory on Flexible Substrates. *Adv. Mater.* **2017**, *29* (11), 1605699. https://doi.org/10.1002/adma.201605699.

(4) Pucher, T.; Puebla, S.; Zamora, V.; Sánchez Viso, E.; Rouco, V.; Leon, C.; Garcia-Hernandez, M.; Santamaria, J.; Munuera, C.; Castellanos-Gomez, A. Strong Electrostatic Control of Excitonic Features in MoS2 by a Free-Standing Ultrahigh-κ Ferroelectric Perovskite. *Adv. Funct. Mater.* **2024**, *34* (52), 2409447. https://doi.org/10.1002/adfm.202409447.

(5) Sen, D.; Ravichandran, H.; Das, M.; Venkatram, P.; Choo, S.; Varshney, S.; Zhang, Z.; Sun, Y.; Shah, J.; Subbulakshmi Radhakrishnan, S.; Saha, A.; Hazra, S.; Chen, C.; Redwing, J. M.; Mkhoyan, K. A.; Gopalan, V.; Yang, Y.; Jalan, B.; Das, S. Multifunctional 2D FETs Exploiting Incipient Ferroelectricity in Freestanding SrTiO3 Nanomembranes at Sub-Ambient Temperatures. *Nat. Commun.* **2024**, *15* (1), 10739. https://doi.org/10.1038/s41467-024-54231-z.

(6) Damodaran, A. R.; Agar, J. C.; Pandya, S.; Chen, Z.; Dedon, L.; Xu, R.; Apgar, B.; Saremi, S.; Martin, L. W. New Modalities of Strain-Control of Ferroelectric Thin Films. *J. Phys. Condens. Matter* **2016**, *28* (26), 263001. https://doi.org/10.1088/0953-8984/28/26/263001.

(7) Feigl, L.; Yudin, P.; Stolichnov, I.; Sluka, T.; Shapovalov, K.; Mtebwa, M.; Sandu, C. S.; Wei, X.-K.; Tagantsev, A. K.; Setter, N. Controlled Stripes of Ultrafine Ferroelectric Domains. *Nat. Commun.* **2014**, *5* (1), 4677. https://doi.org/10.1038/ncomms5677.

(8) Lu, D.; Baek, D. J.; Hong, S. S.; Kourkoutis, L. F.; Hikita, Y.; Hwang, H. Y. Synthesis of Freestanding Single-Crystal Perovskite Films and Heterostructures by Etching of Sacrificial Water-Soluble Layers. *Nat. Mater.* **2016**, *15* (12), 1255–1260. https://doi.org/10.1038/nmat4749.

(9) Chiabrera, F. M.; Yun, S.; Li, Y.; Dahm, R. T.; Zhang, H.; Kirchert, C. K. R.; Christensen, D. V.; Trier, F.; Jespersen, T. S.; Pryds, N. Freestanding Perovskite Oxide Films: Synthesis, Challenges, and Properties. *Ann. Phys.* **2022**, *534* (9), 2200084. https://doi.org/10.1002/andp.202200084.

(10) Bakaul, S. R.; Serrao, C. R.; Lee, M.; Yeung, C. W.; Sarker, A.; Hsu, S.-L.; Yadav, A. K.; Dedon, L.; You, L.; Khan, A. I.; Clarkson, J. D.; Hu, C.; Ramesh, R.; Salahuddin, S. Single Crystal Functional Oxides on Silicon. *Nat. Commun.* **2016**, *7* (1), 10547. https://doi.org/10.1038/ncomms10547.

(11) Huang, J.-K.; Wan, Y.; Shi, J.; Zhang, J.; Wang, Z.; Wang, W.; Yang, N.; Liu, Y.; Lin, C.-H.; Guan, X.; Hu, L.; Yang, Z.-L.; Huang, B.-C.; Chiu, Y.-P.; Yang, J.; Tung, V.; Wang, D.; Kalantar-Zadeh, K.; Wu, T.; Zu, X.; Qiao, L.; Li, L.-J.; Li, S. High-κ Perovskite Membranes as Insulators for Two-Dimensional Transistors. *Nature* **2022**, *605* (7909), 262–267. https://doi.org/10.1038/s41586-022-04588-2.

(12) Zhong, H.; Li, M.; Zhang, Q.; Yang, L.; He, R.; Liu, F.; Liu, Z.; Li, G.; Sun, Q.; Xie, D.; Meng, F.; Li, Q.; He, M.; Guo, E.; Wang, C.; Zhong, Z.; Wang, X.; Gu, L.; Yang, G.; Jin,

K.; Gao, P.; Ge, C. Large-Scale $Hf_{0.5}Zr_{0.5}O_2$ Membranes with Robust Ferroelectricity. *Adv. Mater.* **2022**, *34* (24), 2109889. https://doi.org/10.1002/adma.202109889.

(13) Dong, G.; Li, S.; Yao, M.; Zhou, Z.; Zhang, Y.-Q.; Han, X.; Luo, Z.; Yao, J.; Peng, B.; Hu, Z.; Huang, H.; Jia, T.; Li, J.; Ren, W.; Ye, Z.-G.; Ding, X.; Sun, J.; Nan, C.-W.; Chen, L.-Q.; Li, J.; Liu, M. Super-Elastic Ferroelectric Single-Crystal Membrane with Continuous Electric Dipole Rotation. *Science* **2019**, *366* (6464), 475–479. https://doi.org/10.1126/science.aay7221.

(14) Cai, S.; Lun, Y.; Ji, D.; Lv, P.; Han, L.; Guo, C.; Zang, Y.; Gao, S.; Wei, Y.; Gu, M.; Zhang, C.; Gu, Z.; Wang, X.; Addiego, C.; Fang, D.; Nie, Y.; Hong, J.; Wang, P.; Pan, X. Enhanced Polarization and Abnormal Flexural Deformation in Bent Freestanding Perovskite Oxides. *Nat. Commun.* **2022**, *13* (1), 5116. https://doi.org/10.1038/s41467-022-32519-2.

(15) Guo, Y.; Peng, B.; Lu, G.; Dong, G.; Yang, G.; Chen, B.; Qiu, R.; Liu, H.; Zhang, B.; Yao, Y.; Zhao, Y.; Li, S.; Ding, X.; Sun, J.; Liu, M. Remarkable Flexibility in Freestanding Single-Crystalline Antiferroelectric PbZrO3 Membranes. *Nat. Commun.* **2024**, *15* (1), 4414. https://doi.org/10.1038/s41467-024-47419-w.

(16) Degezelle, A.; Burcea, R.; Gemeiner, P.; Vallet, M.; Dkhil, B.; Fusil, S.; Garcia, V.; Matzen, S.; Lecoeur, P.; Maroutian, T. Strain-Induced Polarization Rotation in Freestanding Ferroelectric Oxide Membranes. *Adv. Electron. Mater.* **2025**, e00266. https://doi.org/10.1002/aelm.202500266.

(17) Elangovan, H.; Barzilay, M.; Seremi, S.; Cohen, N.; Jiang, Y.; Martin, L. W.; Ivry, Y. Giant Superelastic Piezoelectricity in Flexible Ferroelectric $BaTiO_3$ Membranes. *ACS Nano* **2020**, *14* (4), 5053–5060. https://doi.org/10.1021/acsnano.0c01615.

(18) Li, Y.; Zatterin, E.; Conroy, M.; Pylypets, A.; Borodavka, F.; Björling, A.; Groenendijk, D. J.; Lesne, E.; Clancy, A. J.; Hadjimichael, M.; Kepaptsoglou, D.; Ramasse, Q. M.; Caviglia, A. D.; Hlinka, J.; Bangert, U.; Leake, S. J.; Zubko, P. Electrostatically Driven Polarization Flop and Strain-Induced Curvature in Free-Standing Ferroelectric Superlattices. *Adv. Mater.* **2022**, *34* (15), 2106826. https://doi.org/10.1002/adma.202106826.

(19) Wang, Q.; Wang, J.; Fang, H.; Chen, Y.; Han, Y.; Liu, H.; Wang, D.; Zhang, P.; Shi, C.; Guo, J.; He, B.; Zheng, L.; Lü, W. Polarization Evolution in Morphology-Engineered Freestanding Single-Crystalline $BaTiO_3$ Membranes. *J. Phys. Chem. C* **2022**, *126* (38), 16369–16376. https://doi.org/10.1021/acs.jpcc.2c04404.

(20) Pesquera, D.; Cordero-Edwards, K.; Checa, M.; Ivanov, I.; Casals, B.; Rosado, M.; Caicedo, J. M.; Casado-Zueras, L.; Pablo-Navarro, J.; Magén, C.; Santiso, J.; Domingo, N.; Catalan, G.; Sandiumenge, F. Hierarchical Domain Structures in Buckled Ferroelectric Free Sheets. *Acta Mater.* **2025**, *293*, 121080. https://doi.org/10.1016/j.actamat.2025.121080.

(21) Han, L.; Fang, Y.; Zhao, Y.; Zang, Y.; Gu, Z.; Nie, Y.; Pan, X. Giant Uniaxial Strain Ferroelectric Domain Tuning in Freestanding $PbTiO_3$ Films. *Adv. Mater. Interfaces* **2020**, *7* (7), 1901604. https://doi.org/10.1002/admi.201901604.

(22) Hong, S. S.; Gu, M.; Verma, M.; Harbola, V.; Wang, B. Y.; Lu, D.; Vailionis, A.; Hikita, Y.; Pentcheva, R.; Rondinelli, J. M.; Hwang, H. Y. Extreme Tensile Strain States in $La_{0.7}Ca_{0.3}MnO_3$ Membranes. *Science* **2020**, *368* (6486), 71–76. https://doi.org/10.1126/science.aax9753.

(23) Xu, R.; Huang, J.; Barnard, E. S.; Hong, S. S.; Singh, P.; Wong, E. K.; Jansen, T.; Harbola, V.; Xiao, J.; Wang, B. Y.; Crossley, S.; Lu, D.; Liu, S.; Hwang, H. Y. Strain-Induced Room-Temperature Ferroelectricity in SrTiO3 Membranes. *Nat. Commun.* **2020**, *11* (1), 3141. https://doi.org/10.1038/s41467-020-16912-3.

(24) Yang, S.; Khare, K.; Lin, P.-C. Harnessing Surface Wrinkle Patterns in Soft Matter. *Adv. Funct. Mater.* **2010**, *20* (16), 2550–2564. https://doi.org/10.1002/adfm.201000034.
(25) Zhou, Y.; Guo, C.; Dong, G.; Liu, H.; Zhou, Z.; Niu, B.; Wu, D.; Li, T.; Huang, H.; Liu, M.; Min, T. Tip-Induced In-Plane Ferroelectric Superstructure in Zigzag-Wrinkled $BaTiO_3$ Thin Films. *Nano Lett.* **2022**, *22* (7), 2859–2866. https://doi.org/10.1021/acs.nanolett.1c05028.
(26) Dong, G.; Li, S.; Li, T.; Wu, H.; Nan, T.; Wang, X.; Liu, H.; Cheng, Y.; Zhou, Y.; Qu, W.; Zhao, Y.; Peng, B.; Wang, Z.; Hu, Z.; Luo, Z.; Ren, W.; Pennycook, S. J.; Li, J.; Sun, J.; Ye, Z.; Jiang, Z.; Zhou, Z.; Ding, X.; Min, T.; Liu, M. Periodic Wrinkle-Patterned Single-Crystalline Ferroelectric Oxide Membranes with Enhanced Piezoelectricity. *Adv. Mater.* **2020**, *32* (50). https://doi.org/10.1002/adma.202004477.
(27) Long, J.; Wang, T.; Tan, C.; Chen, J.; Zhou, Y.; Lun, Y.; Zhang, Y.; Zhong, X.; Wu, Y.; Song, H.; Ouyang, X.; Hong, J.; Wang, J. Self-Recovery of a Buckling $BaTiO_3$ Ferroelectric Membrane. *ACS Appl. Mater. Interfaces* **2023**, *15* (48), 55984–55990. https://doi.org/10.1021/acsami.3c12730.
(28) Wang, J.; Liu, Z.; Wang, Q.; Nie, F.; Chen, Y.; Tian, G.; Fang, H.; He, B.; Guo, J.; Zheng, L.; Li, C.; Lü, W.; Yan, S. Ultralow Strain-Induced Emergent Polarization Structures in a Flexible Freestanding $BaTiO_3$ Membrane. *Adv. Sci.* **2024**, 2401657. https://doi.org/10.1002/advs.202401657.
(29) Yang, X.; Han, L.; Ning, H.; Xu, S.; Hao, B.; Li, Y.-C.; Li, T.; Gao, Y.; Yan, S.; Li, Y.; Gu, C.; Li, W.; Gu, Z.; Lun, Y.; Shi, Y.; Zhou, J.; Hong, J.; Wang, X.; Wu, D.; Nie, Y. Ultralow-Pressure-Driven Polarization Switching in Ferroelectric Membranes. *Nat. Commun.* **2024**, *15* (1), 9281. https://doi.org/10.1038/s41467-024-53436-6.
(30) Lu, H.; Bark, C.-W.; Esque de los Ojos, D.; Alcala, J.; Eom, C. B.; Catalan, G.; Gruverman, A. Mechanical Writing of Ferroelectric Polarization. *Science* **2012**, *336* (6077), 59–61. https://doi.org/10.1126/science.1218693.
(31) Vats, G.; Ravikant; Schoenherr, P.; Kumar, A.; Seidel, J. Low-Pressure Mechanical Switching of Ferroelectric Domains in $PbZr_{0.48}Ti_{0.52}O_3$. *Adv. Electron. Mater.* **2020**, *6* (10), 2000523. https://doi.org/10.1002/aelm.202000523.
(32) Lu, X.; Chen, Z.; Cao, Y.; Tang, Y.; Xu, R.; Saremi, S.; Zhang, Z.; You, L.; Dong, Y.; Das, S.; Zhang, H.; Zheng, L.; Wu, H.; Lv, W.; Xie, G.; Liu, X.; Li, J.; Chen, L.; Chen, L.-Q.; Cao, W.; Martin, L. W. Mechanical-Force-Induced Non-Local Collective Ferroelastic Switching in Epitaxial Lead-Titanate Thin Films. *Nat. Commun.* **2019**, *10* (1), 3951. https://doi.org/10.1038/s41467-019-11825-2.
(33) Rochon, P.; Batalla, E.; Natansohn, A. Optically Induced Surface Gratings on Azoaromatic Polymer Films. *Appl. Phys. Lett.* **1995**, *66* (2), 136–138. https://doi.org/10.1063/1.113541.
(34) Natansohn, A.; Rochon, P.; Ho, M.-S.; Barrett, C. Azo Polymers for Reversible Optical Storage. 6. Poly[4-[2-(Methacryloyloxy)Ethyl]Azobenzene]. *Macromolecules* **1995**, *28* (12), 4179–4183. https://doi.org/10.1021/ma00116a019.
(35) Yadavalli, N. S.; Korolkov, D.; Moulin, J.-F.; Krutyeva, M.; Santer, S. Probing Opto-Mechanical Stresses within Azobenzene-Containing Photosensitive Polymer Films by a Thin Metal Film Placed Above. *ACS Appl. Mater. Interfaces* **2014**, *6* (14), 11333–11340. https://doi.org/10.1021/am501870t.
(36) Strobelt, J.; Stolz, D.; Leven, M.; Soelen, M. V.; Kurlandski, L.; Abourahma, H.; McGee, D. J. Optical Microstructure Fabrication Using Structured Polarized Illumination. *Opt. Express* **2022**, *30* (5), 7308–7318. https://doi.org/10.1364/OE.451414.
(37) Bakaul, S. R.; Kim, J.; Hong, S.; Cherukara, M. J.; Zhou, T.; Stan, L.; Serrao, C. R.; Salahuddin, S.; Petford-Long, A. K.; Fong, D. D.; Holt, M. V. Ferroelectric Domain Wall

Motion in Freestanding Single-Crystal Complex Oxide Thin Film. *Adv. Mater.* **2020**, *32* (4), 1907036. https://doi.org/10.1002/adma.201907036.
(38) Bourlier, Y.; Bérini, B.; Frégnaux, M.; Fouchet, A.; Aureau, D.; Dumont, Y. Transfer of Epitaxial $SrTiO_3$ Nanothick Layers Using Water-Soluble Sacrificial Perovskite Oxides. *ACS Appl. Mater. Interfaces* **2020**, *12* (7), 8466–8474. https://doi.org/10.1021/acsami.9b21047.
(39) Strobelt, J.; Santer, S.; Abourahma, H.; Music, M.; Farzan, Z.; Nezamis, P.; Leon, R.; McGee, D. J. Photomechanical Azopolymers and Digital Polarization Optics: A Versatile Platform for Surface Microstructure Fabrication. *ACS Appl. Opt. Mater.* **2025**. https://doi.org/10.1021/acsaom.5c00038.
(40) Tverdokhleb, N.; Loebner, S.; Yadav, B.; Santer, S.; Saphiannikova, M. Viscoplastic Modeling of Surface Relief Grating Growth on Isotropic and Preoriented Azopolymer Films. *Polymers* **2023**, *15* (2), 463. https://doi.org/10.3390/polym15020463.
(41) Yadavalli, N. S. Advances in Experimental Methods to Probe Surface Relief Grating Formation Mechanism in Photosensitive Materials. PhD, University of Potsdam **2014**.
(42) Pertsev, N. A.; Zembilgotov, A. G.; Tagantsev, A. K. Effect of Mechanical Boundary Conditions on Phase Diagrams of Epitaxial Ferroelectric Thin Films. *Phys. Rev. Lett.* **1998**, *80* (9), 1988–1991. https://doi.org/10.1103/PhysRevLett.80.1988.
(43) Kukhar, V. G.; Pertsev, N. A.; Kohlstedt, H.; Waser, R. Polarization States of Polydomain Epitaxial $Pb(Zr_{1-x}Ti_x)O_3$ Thin Films and Their Dielectric Properties. *Phys. Rev. B* **2006**, *73* (21), 214103. https://doi.org/10.1103/PhysRevB.73.214103.
(44) Pertsev, N. A.; Tagantsev, A. K.; Setter, N. Phase Transitions and Strain-Induced Ferroelectricity in $SrTiO_3$ Epitaxial Thin Films. *Phys. Rev. B* **2000**, *61* (2), R825–R829. https://doi.org/10.1103/PhysRevB.61.R825.
(45) Barbosa, J. G.; Gomes, I. T.; Pereira, M. R.; Moura, C.; Mendes, J. A.; Almeida, B. G. Structural and Dielectric Properties of Laser Ablated BaTiO3 Films Deposited over Electrophoretically Dispersed CoFe2O4 Grains. *J. Appl. Phys.* **2014**, *116* (16), 164112. https://doi.org/10.1063/1.4900516.
(46) Guo, R.; You, L.; Lin, W.; Abdelsamie, A.; Shu, X.; Zhou, G.; Chen, S.; Liu, L.; Yan, X.; Wang, J.; Chen, J. Continuously Controllable Photoconductance in Freestanding BiFeO3 by the Macroscopic Flexoelectric Effect. *Nat. Commun.* **2020**, *11* (1), 2571. https://doi.org/10.1038/s41467-020-16465-5.
(47) Shi, Q.; Parsonnet, E.; Cheng, X.; Fedorova, N.; Peng, R.-C.; Fernandez, A.; Qualls, A.; Huang, X.; Chang, X.; Zhang, H.; Pesquera, D.; Das, S.; Nikonov, D.; Young, I.; Chen, L.-Q.; Martin, L. W.; Huang, Y.-L.; Íñiguez, J.; Ramesh, R. The Role of Lattice Dynamics in Ferroelectric Switching. *Nat. Commun.* **2022**, *13* (1), 1110. https://doi.org/10.1038/s41467-022-28622-z.
(48) Zubko, P.; Catalan, G.; Tagantsev, A. K. Flexoelectric Effect in Solids. *Annu. Rev. Mater. Res.* **2013**, *43* (1), 387–421. https://doi.org/10.1146/annurev-matsci-071312-121634.
(49) Tian, D.; Jeong, D.-Y.; Fu, Z.; Chu, B. Flexoelectric Effect of Ferroelectric Materials and Its Applications. *Actuators* **2023**, *12* (3), 114. https://doi.org/10.3390/act12030114.
(50) Ma, W.; Cross, L. E. Flexoelectric Effect in Ceramic Lead Zirconate Titanate. *Appl. Phys. Lett.* **2005**, *86* (7), 072905. https://doi.org/10.1063/1.1868078.
(51) Ma, W.; Cross, L. E. Flexoelectricity of Barium Titanate. *Appl. Phys. Lett.* **2006**, *88* (23). https://doi.org/10.1063/1.2211309.
(52) Lyu, L.; Song, C.; Wang, Y.; Wu, D.; Zhang, Y.; Su, S.; Huang, B.; Li, C.; Xu, M.; Li, J. Anomalously High Apparent Young's Modulus of Ultrathin Freestanding PZT Films Revealed by Machine Learning Empowered Nanoindentation. *Adv. Mater.* **2025**, *37* (1), 2412635. https://doi.org/10.1002/adma.202412635.

(53) Segantini, G.; Tovaglieri, L.; Roh, C. J.; Hsu, C.-Y.; Cho, S.; Bulanadi, R.; Ondrejkovic, P.; Marton, P.; Hlinka, J.; Gariglio, S.; Alexander, D. T. L.; Paruch, P.; Triscone, J.-M.; Lichtensteiger, C.; Caviglia, A. D. Curvature-Controlled Polarization in Adaptive Ferroelectric Membranes. *Small* *n/a* (n/a), e06338. https://doi.org/10.1002/smll.202506338.

(54) Cherifi-Hertel, S.; Bulou, H.; Hertel, R.; Taupier, G.; Dorkenoo, K. D. (Honorat); Andreas, C.; Guyonnet, J.; Gaponenko, I.; Gallo, K.; Paruch, P. Non-Ising and Chiral Ferroelectric Domain Walls Revealed by Nonlinear Optical Microscopy. *Nat. Commun.* **2017**, *8* (1), 15768. https://doi.org/10.1038/ncomms15768.

(55) Gu, Y.; Li, M.; Morozovska, A. N.; Wang, Y.; Eliseev, E. A.; Gopalan, V.; Chen, L.-Q. Flexoelectricity and Ferroelectric Domain Wall Structures: Phase-Field Modeling and DFT Calculations. *Phys. Rev. B* **2014**, *89* (17), 174111. https://doi.org/10.1103/PhysRevB.89.174111.

(56) Pertsev, N. A.; Zembilgotov, A. G.; Tagantsev, A. K. Equilibrium States and Phase Transitions in Epitaxial Ferroelectric Thin Films. *Ferroelectrics* **1999**, *223* (1), 79–90. https://doi.org/10.1080/00150199908260556.

(57) Lee, J. W.; Eom, K.; Paudel, T. R.; Wang, B.; Lu, H.; Huyan, H. X.; Lindemann, S.; Ryu, S.; Lee, H.; Kim, T. H.; Yuan, Y.; Zorn, J. A.; Lei, S.; Gao, W. P.; Tybell, T.; Gopalan, V.; Pan, X. Q.; Gruverman, A.; Chen, L. Q.; Tsymbal, E. Y.; Eom, C. B. In-Plane Quasi-Single-Domain BaTiO3 via Interfacial Symmetry Engineering. *Nat. Commun.* **2021**, *12* (1), 6784. https://doi.org/10.1038/s41467-021-26660-7.

(58) Horcas, I.; Fernández, R.; Gómez-Rodríguez, J. M.; Colchero, J.; Gómez-Herrero, J.; Baro, A. M. WSXM: A Software for Scanning Probe Microscopy and a Tool for Nanotechnology. *Rev. Sci. Instrum.* **2007**, *78* (1), 013705. https://doi.org/10.1063/1.2432410.

(59) Jelken, J.; Santer, S. Light Induced Reversible Structuring of Photosensitive Polymer Films. *RSC Adv.* **2019**, *9* (35), 20295–20305. https://doi.org/10.1039/C9RA02571E.

## Supplementary information

# Light-Induced Photomechanical Patterning of Ferroelectric Polarization

*Alban Degezelle[1*], Jonas Strobelt[2], Sarah Loebner[2], Moussa Mebarki[3], Stéphane Fusil[4], Vincent Garcia[4], Bruno Bérini[3], Vincent Polewczyk[3], Yves Dumont[3], Sylvia Matzen[1], Philippe Lecoeur[1], Svetlana Santer[2], Thomas Maroutian[1†]*

[* alban.degezelle@universite-paris-saclay.fr](mailto:alban.degezelle@universite-paris-saclay.fr)

[†][thomas.maroutian@cnrs.fr](mailto:thomas.maroutian@cnrs.fr)

[1] Centre de Nanosciences et de Nanotechnologies, CNRS, Université Paris-Saclay, 91120 Palaiseau, France

[2] Institute of Physics and Astronomy, University of Potsdam, 14476 Potsdam, Germany

[3] Université Paris-Saclay, UVSQ, CNRS, GEMaC, 78000 Versailles, France

[4] Laboratoire Albert Fert, CNRS, Thales, Université Paris-Saclay, 91767 Palaiseau, France

## 1. Mechanical modelling of bending strain

To compute the in-plane strain $\varepsilon_{xx}$ along the topography profile direction (here illustrated along the $x$ axis), we first calculate the neutral layer position relative to the oxide/electrode interface, defined at $z = 0$ ($z$ thickness direction). The membrane top surface is then at $z = +t_M$ and the electrode bottom surface at $z = -t_{E_l}$.

The position δ along the $z$-axis of the neutral layer is given by the following relation:

$$\delta = \frac{1}{2} \times \frac{(t_M^2 E_M - t_E^2 E_{El})}{(t_M E_M + t_{El} E_{El})} \quad \text{(S1)}$$

With $E_M$ and $E_{El}$ the Young modulus of the membrane and its electrode, respectively.

The strain map computation is done by using an 8-bit image of the topography. We then decompose the 2D image as a sum of $y$-1D profile along each $x$-line, converting the grey scale values to the corresponding topographic height values:

$$0 \rightarrow H_{min} \; ; \; 255 \rightarrow H_{max} \quad \text{(S2)}$$

For smoother strain map visualization, we applied a soft 1D gaussian blur G(x) on the profile w(x), as follows:

$$G(x) = \frac{1}{\sqrt{2\pi}} e^{-x^2/(2\sigma^2)} \quad \text{(S3)}$$

$$\widetilde{W}(\mathrm{x}) = \int_{-\infty}^{\infty} w(x') G(x - x') dx' \quad \text{(S4)}$$

We then compute the strain, line by line along the $x$-axis using the 1D Euler-Bernoulli formula for a symmetrical bending strain (S5). Gaussian blur was maintained between $\sigma = 1 - 3 \ (px)$, to be visually pleasing without altering too much the computed strain value.

$$\varepsilon_{xx}(x, \delta) = -Z \frac{d^2 \widetilde{W}(x)}{dx^2} \quad \text{(S5)}$$

Equation (S5) describes the in-plane bending strain along $x$, computed at the top surface for $Z = t_M - \delta$. Therefore, the relevant in-plane strain gradient $\varepsilon_{xx,z}$ (across the film thickness) is given by:

$$\varepsilon_{xx,z} = -\frac{d^2\widetilde{W}(x)}{dx^2} \qquad \text{(S6)}$$

This gradient is positive in tensile strain region and negative in compressive strain region. With a positive flexoelectric coefficient $\mu_{zxxz}$ (transverse flexoelectric coefficient, also called $\mu_{12}$in the literature), the flexoelectric polarization is pointing upward in tensile region and downward in compressive region and is expressed by:

$$P_{Flexo} = \mu_{zxxz}\varepsilon_{xx,z} \qquad \text{(S7)}$$

**Table T1: Reported Young's modulus (E) values for relevant oxide materials used in the neutral layer computation.**

| **Material** | **E (GPa)** |
|---|---|
| **PTO** | 146 [1], 153 [2] |
| **LSMO** | 127 [3] |
| **STO** | 225[4] |
| **SRO** | 45[5] |
| **BTO** | 139 [6] |

**Table S1:** Material Young's modulus from literature**.**

We want to point out that recent studies show that the Young modulus of perovskite oxides membranes is highly prone to change when they are thinned or subjected to high bending. Notably, studies were done for STO [7], PZT[8] and SRO[5].

## 2. PAZO deflection for hill-and-valley patterns

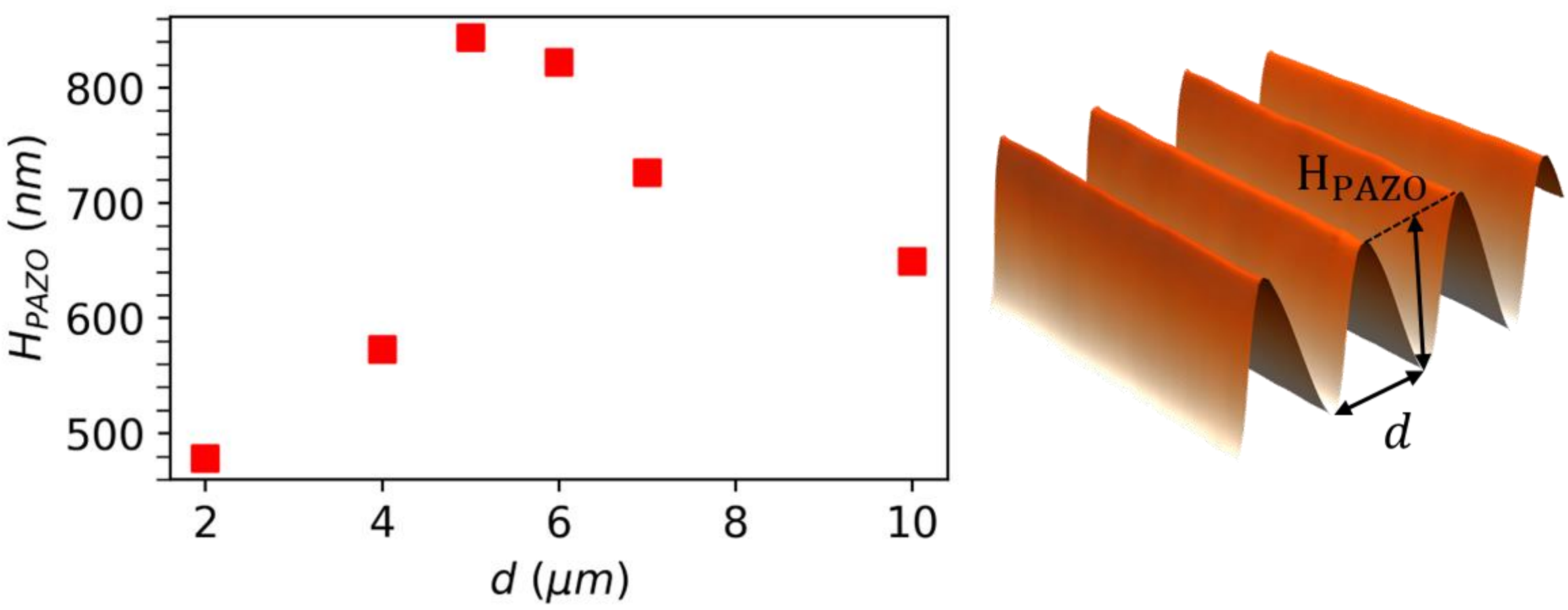

**Figure S1:** Deformation height for various H&V patterns on PAZO alone. Measured by AFM. The period is denoted $d$ and the height $H_{PAZO}$.

## 3. STO hill-and-valley pattern

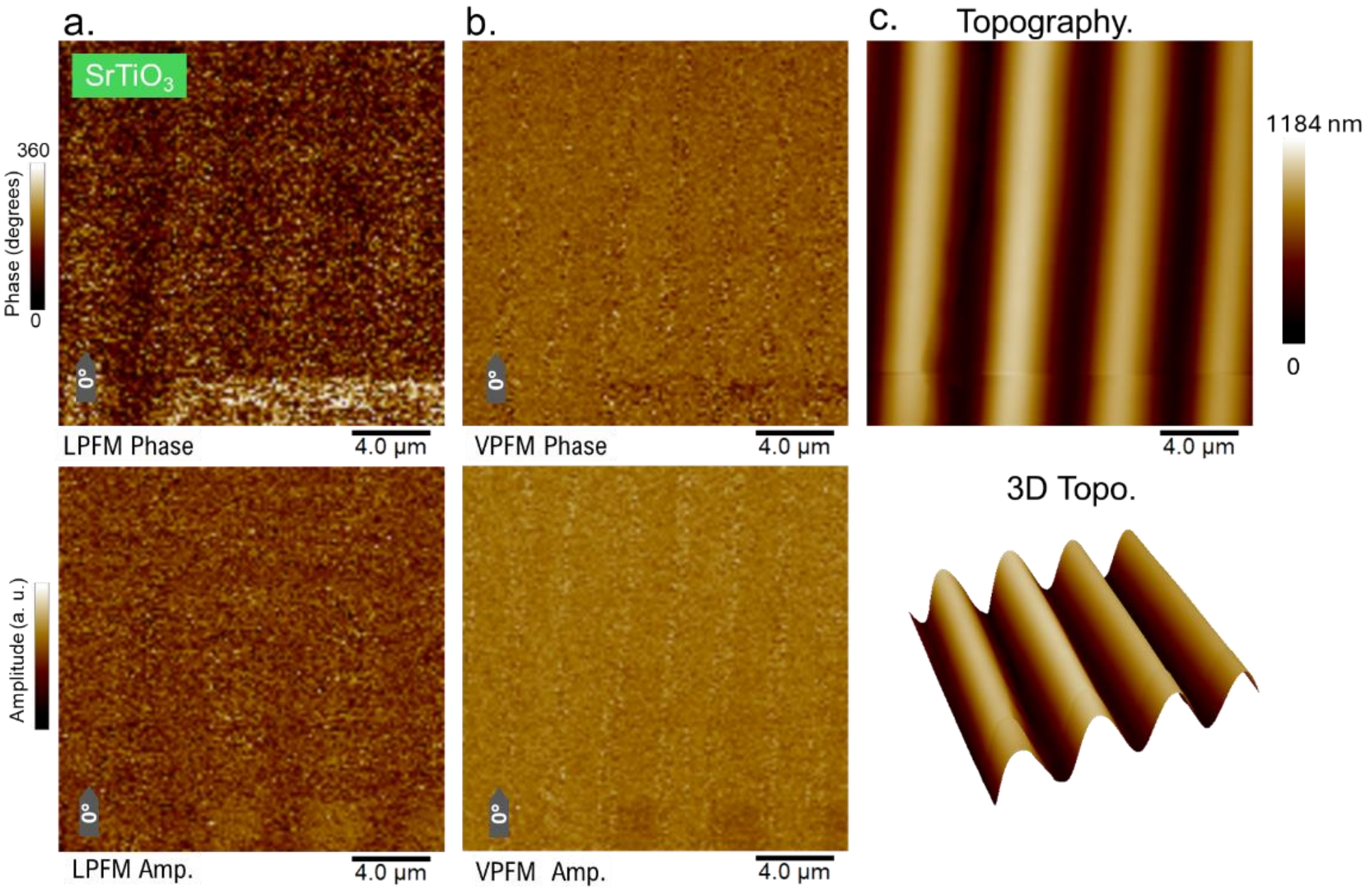

**Figure S2:** LPFM (a) and VPFM (b) scans on STO H&V pattern. No net polarization component is observed due to insufficient strain/gradient.

## 4. XRD scans of the as-grown samples

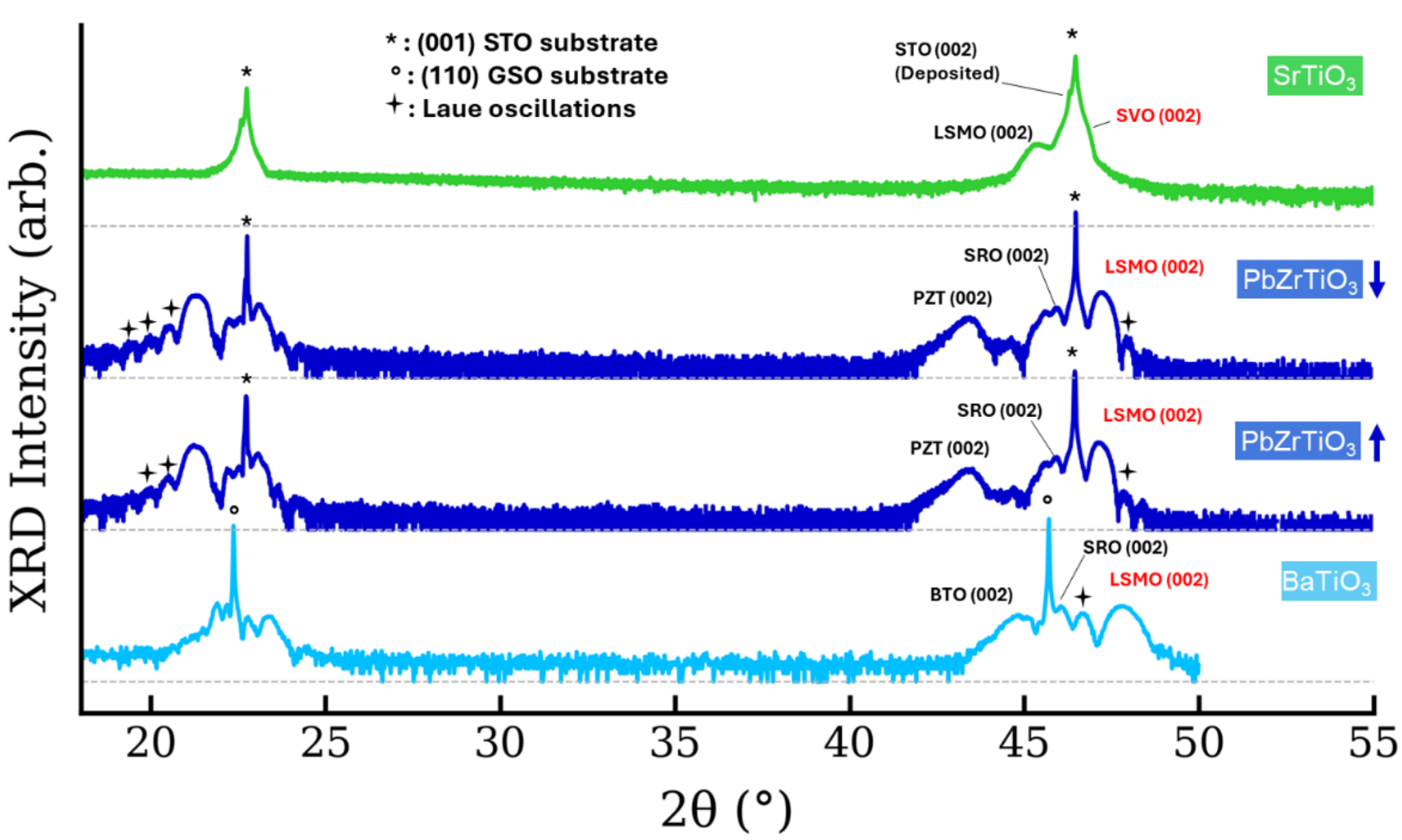

**Figure S3:** XRD θ-2θ scans of typical PZT (up and down polarization, dark blue), BTO (light blue) and STO (green) as-grown stacks with their respective electrodes, sacrificial layers (red) and substrates.

## 5. VPFM DC box-in-box writing scans of the as-transferred membranes

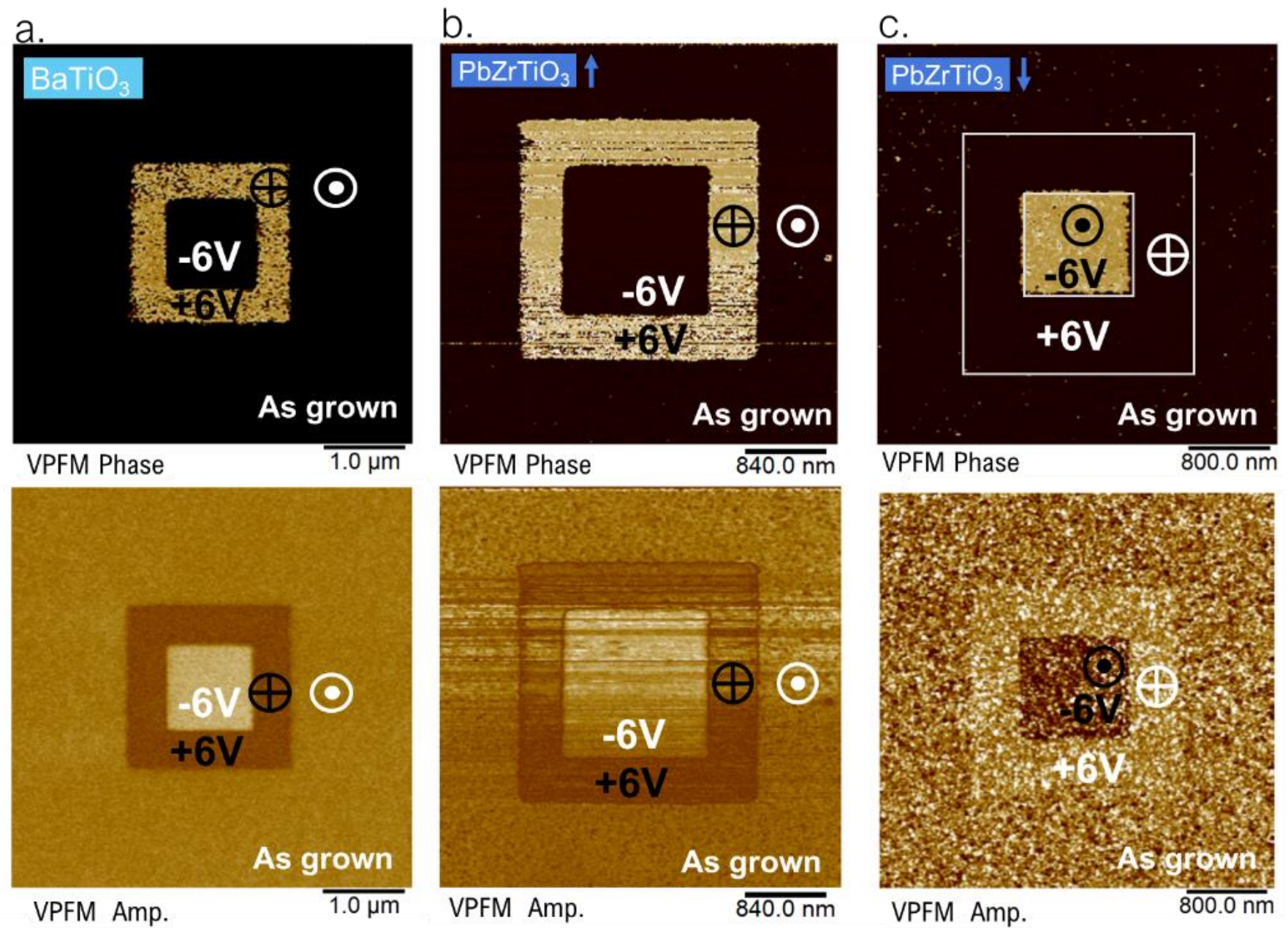

**Figure S4:** Box-in-box DC writing of BTO (a), PZT up (b) and PZT down (c) membranes on PAZO in a flat area, with voltage bias applied to the PFM tip and bottom electrode grounded. The ferroelectric films retain their as-grown polarization orientation upon transfer to the PAZO polymer.

## 6. Lower deflection BTO hill-and-valley pattern

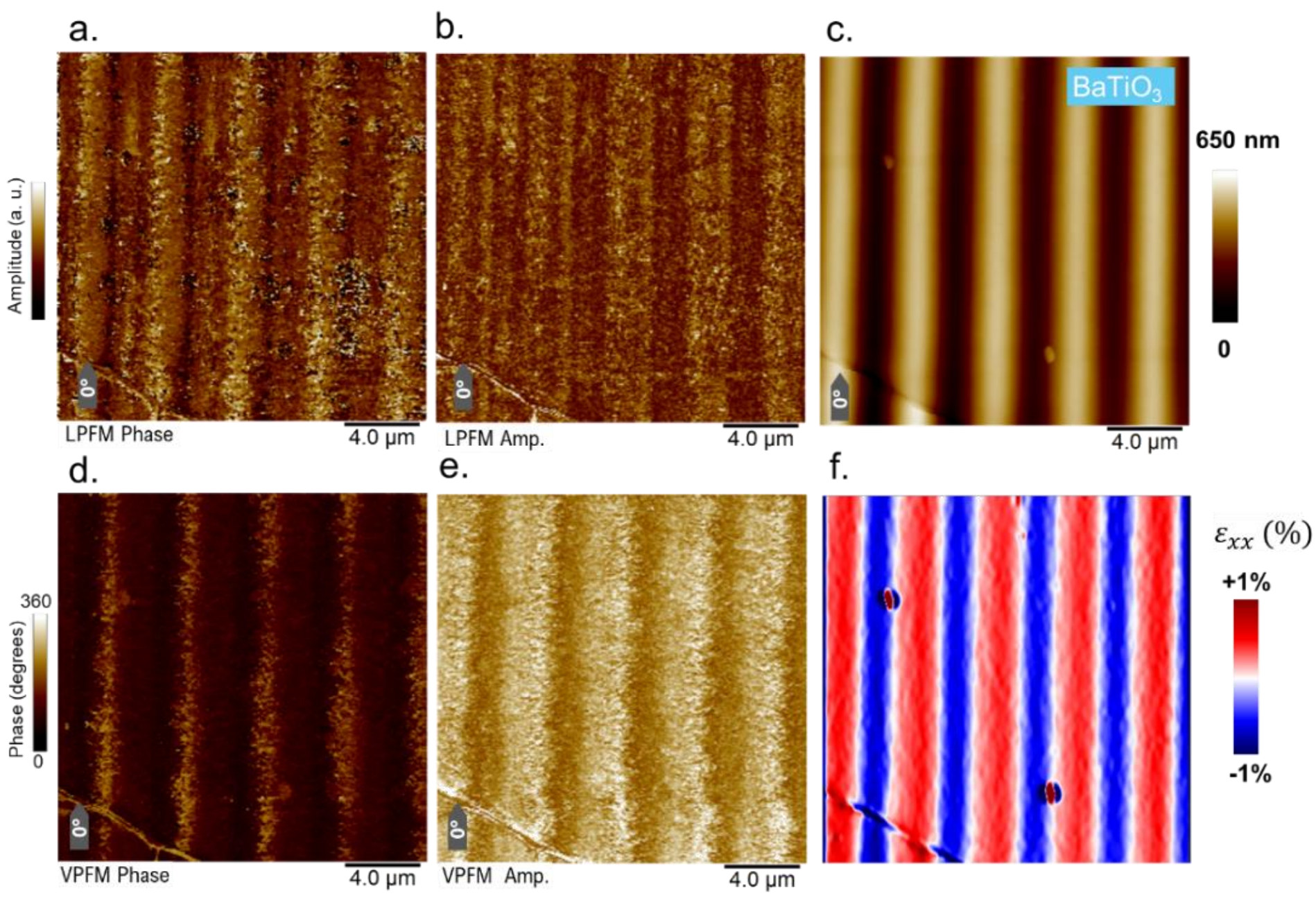

**Figure S5:** PFM scan of a 4 µm-period H&V pattern in BTO. LPFM phase (a) and amplitude (b); VPFM phase (d) and amplitude (e); Topography (c) and its associated strain map (f).

Within the same BTO membrane, we investigated a H&V grating, this time with a 4 µm period instead of 6 µm in the main article. As anticipated, this pattern exhibits a reduced deflection height of 650 nm. The LPFM scan of this HV pattern (**Figure S5a-b**) reveals no 90° polarization rotation at the hilltops (Maximum $\varepsilon_{xx} \approx 0.45\%$), in contrast to the pattern with higher deflection. Strikingly, the VPFM scan (**Figure S5d–e**) again demonstrates a substantial reduction in VPFM amplitude. However, unlike the previous case with higher deflection (**Figure 3**), no subsequent amplitude increase is observed anywhere along the valleys. While the VPFM amplitude decrease remains coupled with a phase shift, it does not culminate in a full reversal, unlike the higher-deflection HV pattern depicted in **Figure 3b**.

In the case of this lower deflection (650 nm maximum), the average maximum strain gradient in the valley decreases to approximately $-4 \times 10^5$ m$^{-1}$ and never locally attains values as high as $10^6$ m$^{-1}$. While the resulting flexoelectric polarization remains sufficient to induce a rotation of the total polarization, it is insufficient to achieve a full 180° reversal. This is evidenced by the absence of a VPFM amplitude enhancement at the valley depth (**Figure S6a**, top panel) and the gradual evolution of the VPFM phase across the valleys (**Figure S6a**, middle panel).

This is coherent with **Figure 5** in the main text showing that the flexoelectric polarization should overcomes the polarization when the gradient reaches around $5 \times 10^5$ m$^{-1}$.

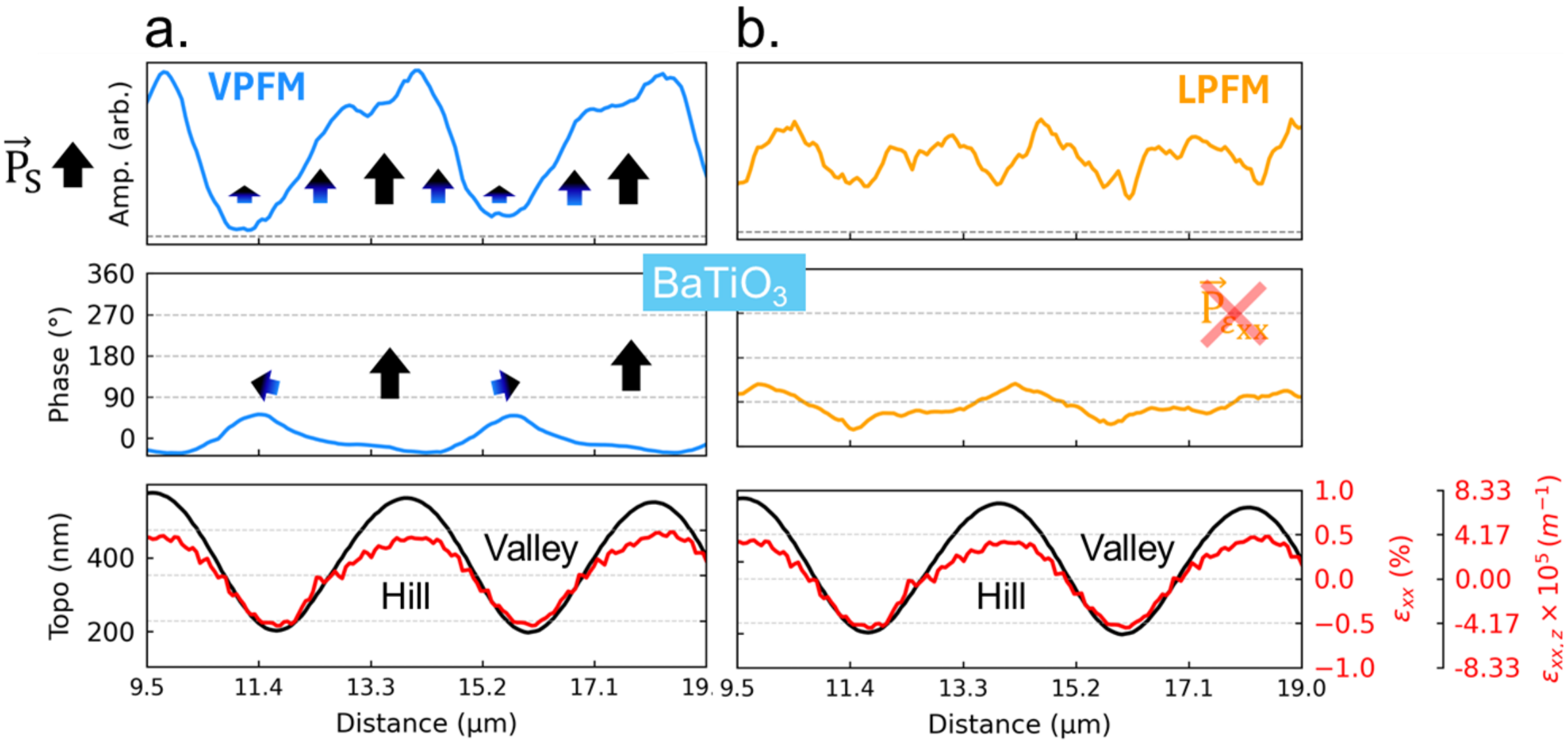

**Figure S6:** (a) VPFM signal (blue line) in a BTO 4 $\mu m$ period HV pattern with amplitude (top panel) and phase (middle panel). Arrows depict the flexoelectric component reducing the total polarization by counteracting the spontaneous upward one ($P_S$, in black). The height profile is shown in the bottom panel (black line), together with the computed in-plane bending strain (red line). Note that the in-plane strain gradient has the exact same curve, albeit with a separate scale as indicated on the bottom right. (b) Corresponding LPFM signal (orange line) with amplitude (top panel) and phase (middle panel), showing no clear LPFM amplitude or phase change. No IP polarized region ($P_{\varepsilon_{xx}}$, in orange) is observed (red crossed). All PFM profiles were extracted from the data shown in **Figure S5** and averaged in the wrinkle length direction.

## 7. PZT hill-and-valley pattern

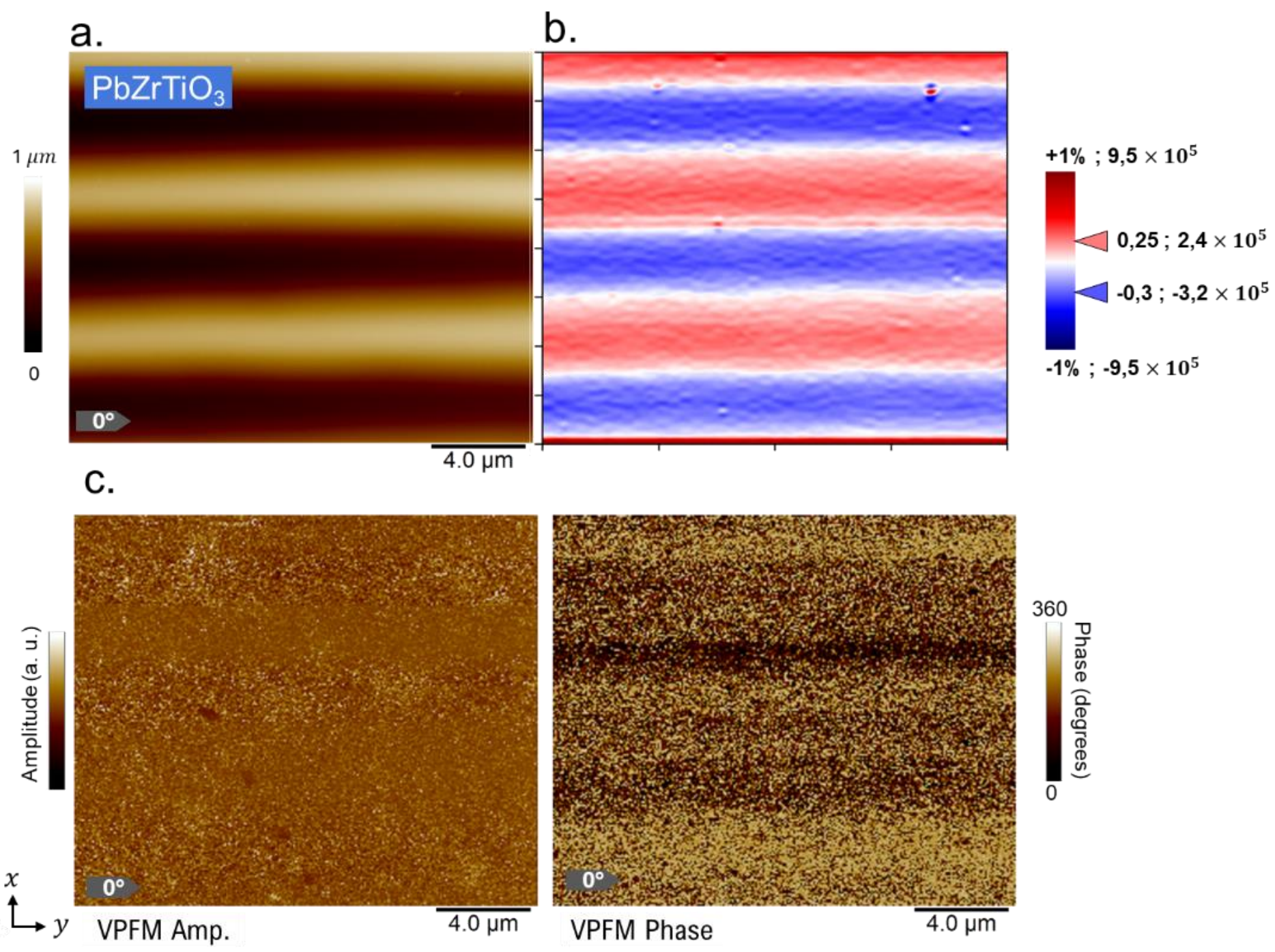

**Figure S7:** (a) Topography, (b) Strain map and (c) VPFM scans of a down polarization (-c) PZT thin membrane. We observe that despite high gradient values, no flexoelectric switch is observed in the VPFM amplitude and phase. The pattern period is 6 $\mu m$ and the maximum deflection 900 nm.

## 8. Zoomed PFM scans

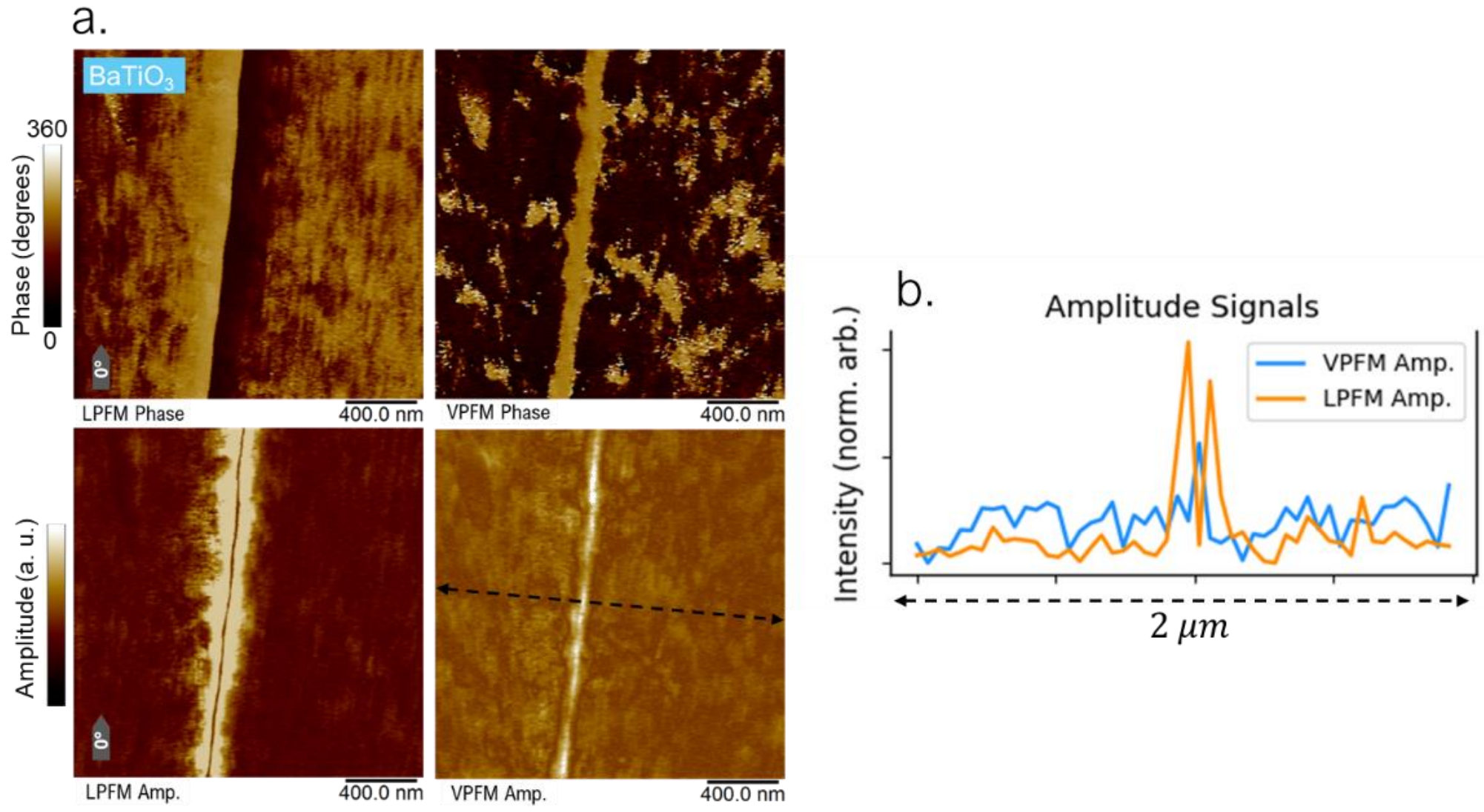

**Figure S8:** (a) LPFM and VPFM scans on top of the BTO hill where the strain induced polarization switch occurs. We observe a bright (downward) VPFM phase at the location of the IP domain wall barrier. A VPFM amplitude spike is also visible at this exact location as evidenced by the cross-section plot of both PFM amplitude channels in (b). Dashed line shows the cross-section profile location.

## References

(1) Li, Z.; Grimsditch, M.; Foster, C. M.; Chan, S.-K. Dielectric and Elastic Properties of Ferroelectric Materials at Elevated Temperature. *J. Phys. Chem. Solids* **1996**, *57* (10), 1433–1438. https://doi.org/10.1016/0022-3697(96)00009-1.

(2) Long, J.; Yang, L.; Wei, X. Lattice, Elastic Properties and Debye Temperatures of ATiO3 (A = Ba, Ca, Pb, Sr) from First-Principles. *J. Alloys Compd.* **2013**, *549*, 336–340. https://doi.org/10.1016/j.jallcom.2012.08.120.

(3) Manca, N.; Remaggi, F.; Plaza, A. E.; Varbaro, L.; Bernini, C.; Pellegrino, L.; Marré, D. Stress Analysis and Q-Factor of Free-Standing (La,Sr)MnO3 Oxide Resonators. *Small* **2022**, *18* (35), 2202768. https://doi.org/10.1002/smll.202202768.

(4) Bernard, O.; Andrieux, M.; Poissonnet, S.; Huntz, A.-M. Mechanical Behaviour of Ferroelectric Films on Perovskite Substrate. *J. Eur. Ceram. Soc.* **2004**, *24* (5), 763–773. https://doi.org/10.1016/S0955-2219(03)00324-8.

(5) Su, P.; Wen, H.; Zhang, Y.; Tan, C.; Zhong, X.; Wu, Y.; Song, H.; Zhou, Y.; Li, Y.; Liu, M.; Wang, J. Super-Flexibility in Freestanding Single-Crystal SrRuO3 Conductive Oxide Membranes. *ACS Appl. Electron. Mater.* **2022**, *4* (6), 2987–2992. https://doi.org/10.1021/acsaelm.2c00422.

(6) Park, Y.-B.; Dicken, M. J.; Xu, Z.-H.; Li, X. Nanoindentation of the a and c Domains in a Tetragonal BaTiO3 Single Crystal. *J. Appl. Phys.* **2007**, *102* (8), 083507. https://doi.org/10.1063/1.2795664.
(7) Harbola, V.; Crossley, S.; Hong, S. S.; Lu, D.; Birkhölzer, Y. A.; Hikita, Y.; Hwang, H. Y. Strain Gradient Elasticity in $SrTiO_3$ Membranes: Bending versus Stretching. *Nano Lett.* **2021**, *21* (6), 2470–2475. https://doi.org/10.1021/acs.nanolett.0c04787.
(8) Lyu, L.; Song, C.; Wang, Y.; Wu, D.; Zhang, Y.; Su, S.; Huang, B.; Li, C.; Xu, M.; Li, J. Anomalously High Apparent Young's Modulus of Ultrathin Freestanding PZT Films Revealed by Machine Learning Empowered Nanoindentation. *Adv. Mater.* **2025**, *37* (1), 2412635. https://doi.org/10.1002/adma.202412635.